\documentclass[aps,pra,twocolumn,showpacs,groupedaddress,notitlepage,superscriptaddress,floatfix,nofootinbib]{revtex4-2}

\usepackage{amsmath}
\usepackage{amssymb}
\usepackage{amsthm}
\usepackage{braket}
\usepackage{graphicx}
\usepackage{subcaption}
\DeclareMathOperator{\Tr}{Tr}
\DeclareMathOperator{\Var}{Var}

\DeclareMathOperator{\Cov}{Cov}

\usepackage[usenames,dvipsnames,svgnames,table]{xcolor}

\definecolor{blue-violet}{rgb}{0.54, 0.17, 0.89}

\usepackage{hyperref}
\hypersetup{colorlinks=true, linkcolor=RubineRed,
            citecolor=blue-violet, urlcolor=Cerulean}
\usepackage{cleveref}

\usepackage{booktabs}
\usepackage{refcount}

\begin{document}

% --- Title block (place in preamble, after your packages) ---
\title{Experimental Observation of Error-Resilience Phase Transition on Superconducting Quantum Processor}

\author{Emanuel Dallas}
\affiliation{Department of Physics \& Astronomy, University of Southern California, Los Angeles, California 90089, USA}
\affiliation{Center for Quantum Information Science \& Technology, University of Southern California, Los Angeles, California 90089, USA}
\email{dallas@usc.edu}
\author{Mario Morford-Oberst}
\affiliation{Center for Quantum Information Science \& Technology, University of Southern California, Los Angeles, California 90089, USA}
\affiliation{Department of Electrical \& Computer Engineering, University of Southern California, Los Angeles, California 90089, USA}
\author{Eric Weine}
\affiliation{Department of Electrical Engineering and Computer Science, Massachusetts Institute of Technology, Cambridge, MA 02139, USA}

\date{\today}

\begin{abstract}
Random unitaries scramble quantum information into highly nonlocal degrees of freedom, thereby providing protection from local noise. However, as demonstrated in~\cite{enc_dec_original, enc_dec_magic}, this protection may fail once local noise grows sufficiently strong. These works describe a phase transition in a noisy random encoding-decoding circuit model between an error-protecting phase, in which encoded information is recoverable, and an error-vulnerable phase, in which encoded information is lost, with the transition occurring at a critical coherent noise angle $\alpha_c \approx 1.14$. Here, we present the first experimental observation of this transition, conducted on IBM's \texttt{ibm\_pittsburgh} Heron r3 device for system sizes of up to $N = 26$ qubits. We detect this transition via three order parameters: fidelity, R\'enyi-2 entanglement entropy, and stabilizer R\'enyi-2 entropy. For the latter two, we employ randomized measurement protocols to construct estimators, and for all three, we rely on zero-noise extrapolation and M3 SPAM error mitigation. We then jointly model the finite-size scaling collapse across all three order parameters with a Gaussian process regression, yielding a critical angle $\alpha_c \approx 1.01$, well below the theoretical value. Inputting noiseless simulated data into the same model yields $\alpha_c \approx 1.05$, implying that the discrepancy between experiment and theory is due to our encoding circuits, which are not deep enough to be truly Haar-random. Finally, we observe that the same model cannot adequately identify critical behavior in unmitigated, raw experimental data. This indicates that error mitigation techniques are crucial for detecting the transition on hardware.
\end{abstract}

\maketitle

\section{Introduction}

The interaction between unitary dynamics and environmental noise is a rich and complex topic of recent exploration in quantum information theory. On one hand, noise can dissipate information out of open quantum systems and undermine computation. On the other hand, noise can increase the complexity of quantum states~\cite{noise_qc}, likewise increasing the quantity of uniquely quantum resources like entanglement and non-stabilizerness~\cite{SRE,noise_entanglement_markovian, noise_entanglement_monitored, noise_magic_nonunital}.

Randomization is a powerful way to mitigate the deleterious effects of noise. Random unitaries drawn from the Haar measure are typically highly \textit{scrambling}~\cite{haar_chaos_scramble, haar_hayden_preskill, haar_fast_scramblers}, spreading initially localized information throughout nonlocal degrees of freedom. Distributing information nonlocally can increase its robustness to local noise, as the nonlocal correlations that store information persist in spite of these perturbations; this is the idea underlying quantum error correction~\cite{shor_scheme_1995,Steane:96a,Lidar-Brun:book}.

In~\cite{enc_dec_original,enc_dec_magic}, Turkeshi and Sierant study the interaction between random dynamics and noise. They consider the final ``logical" state of a system of qubits, some ``logical" and some ``ancillary," undergoing a Haar-random encoding unitary, followed by a layer of identical single-qubit (coherent) noise on each qubit, then decoding by the encoding's inverse, and finally post-selecting on the ancilla state (see \cref{theory}). This system exhibits a phase transition at a critical noise strength, expressed in the order parameters of logical fidelity with the initial state, entanglement entropy, and non-stabilizerness. Their results highlight two salient features of the interplay between noise and generic unitary dynamics. First, while scrambling protects information, the transition demonstrates that the relationship between noise strength and information protection is highly nontrivial, indicating a regime of protection. Second, the dual nature of noise is directly manifest. It completely destroys the fidelity past the critical point while generating extensive entanglement and non-stabilizerness in the final logical state.

Until now, these results have not been observed on existing quantum hardware. Direct experimental observation poses serious implementation challenges and also requires substantial deviations from the idealized model. In practice, while current devices can execute deep circuits, truly Haar-random circuits require exponentially many two-qubit gates~\cite{haar_exponentially_hard_knill}. Coherent gate errors prevent the encoding and decoding unitaries from being exact inverses, state preparation and measurement (SPAM) errors prevent measurement results from displaying the expected outcomes, and depolarization contracts states toward the maximally mixed state. While the results of~\cite{enc_dec_original, enc_dec_magic} are robust against some types of errors, it is not obvious how an implementation of their model behaves in the presence of correlated and site-dependent noise of current devices. Furthermore, precisely measuring entanglement and non-stabilizerness on NISQ-era devices is difficult and requires protocols with significant overhead, such as randomized measurements~\cite{RM_toolbox, RM_toolbox_entanglement, RM_probing_entanglement, SRE, meas_magic}, which we implement in this work.

In spite of these difficulties, we observe a phase transition in our implementation of the encoding-decoding model as a function of noise angle $\alpha$. We rely on gate error mitigation using zero-noise extrapolation (ZNE) and SPAM error mitigation using matrix-free measurement error mitigation (M3), which require additional circuit overhead, to enhance the observable signal in our data. We implement randomized protocols for measuring entanglement and non-stabilizerness, incorporating our experimental fidelity data in an importance sampling variance-reduction protocol for our entanglement entropy estimator. We then model the phase transition on our experimental data with Gaussian process regression. This yields posterior distributions for the model’s critical parameters across all three order parameters. The posteriors both validate the presence of a transition and allow us to characterize the deviations of real-device behavior from the theoretical model.

In \cref{theory} we review the theoretical model and results of~\cite{enc_dec_original} and~\cite{enc_dec_magic}. In \cref{experiment_implement} we describe our experimental implementation of the model, and we describe the error mitigation and measurement protocols in \cref{error_mitigation} and \cref{measurements}, respectively. We display the experimental data in \cref{experimental_data} and analyze it in \cref{GP_regression}, discussing the results in \cref{GP_regression_results}.

\section{Theoretical Model}\label{theory}

\begin{figure*}[t]
    \centering
    \includegraphics[width=0.6\textwidth]{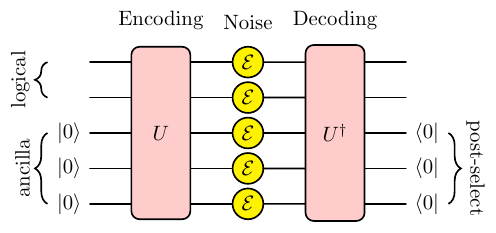} 
    \caption{Circuit diagram of the theoretical model. Ancilla qubits are prepared in the state $\ket{0}$. A global Haar-random encoding $U$ is applied to all qubits. Identical single-qubit noise then acts on each qubit, followed by decoding with $U^\dagger$ and post-selection on ancilla qubits being in their initial state.}
    \label{fig:theory_circuit_diagram}
\end{figure*}

In~\cite{enc_dec_original, enc_dec_magic}, the authors consider a toy model of an encoded quantum memory protocol, depicted in \cref{fig:theory_circuit_diagram}. The system consists of $N$ qubits, of which $k$ form the ``logical" subsystem, denoted by $X$, and the remaining $N-k$ are ancillas, denoted by $\overline{X}$. The initial state $\rho_0$ is a product over this bipartition, i.e., $\rho_0  = \ket{\psi_X}\bra{\psi_X} \otimes \ket{\psi_{\overline{X}}}\bra{{\psi_{\overline{X}}}}$. All $N$ qubits are acted on by an encoding unitary $U$, selected randomly from the Haar measure, followed by a layer of identical local errors $\mathcal{E} = \mathcal{E}_1 \otimes ... \otimes \mathcal{E}_N $ acting on each qubit, and a decoding unitary $U^\dagger$. Then, a measurement is performed to ``check" whether the ancilla qubits are in their initial state. The final (pure) logical state $\rho_X$ is obtained by projecting out the ancilla qubits post-selected on their being in the initial state. We consider the case of coherent noise, in which $\mathcal{E}_i(\rho) = K_i\rho K_i^\dagger = e^{-i\alpha \vec{n}\cdot \vec{\sigma}/2}\rho e^{i\alpha \vec{n}\cdot \vec{\sigma}/2}$, where $\vec{n}$ is a Bloch vector and $\vec{\sigma} \equiv \left\{X, Y, Z \right\}$.

Sampling from the Haar measure ensures that the system lacks a ``preferred" basis. This means that the choices of initial ancilla state and orientation of the coherent-error rotation axis do not affect the theoretical results. For simplicity, let $\ket{\psi_{\overline{X}}} = \ket{0_{\overline{X}}} \equiv \ket{0}^{\otimes (N-k)}$, and let $\vec{n} = \left(0,0,1 \right)$ so that $\forall i, \ K_i = e^{-i\alpha \sigma_Z/2}$. In the following results and in experiment, we also let the initial logical state be $\ket{\psi_X} = \ket{0_X} \equiv \ket{0}^{\otimes k}$.

The authors analyze three properties of the final state $\rho_X$. First, the fidelity between the final and initial logical states: $F \equiv \bra{0_X} \rho_{X} \ket{0_X}$. Second, the entanglement entropy of $\rho_X$, quantified by the qth-order R\'enyi entropy

\begin{equation}
    S_q(\rho_X) = \frac{1}{1-q} \log_2\left(\Tr_{X_1}\left[\left(\Tr_{X_2}(\rho_X)\right)^q \right] \right),
\end{equation}
where $\left(X_1, X_2\right)$ is a qubit bipartition of $X$.

In this work, we focus on the R\'enyi-2 entanglement entropy $S_2(\rho_X) = -\log_2\left(\Tr_{X_1}\left[\left(\Tr_{X_2}(\rho_X)\right)^2 \right] \right)$. Finally, the non-stabilizerness of $\rho_X$, quantified by the qth-order stabilizer R\'enyi entropy~\cite{SRE}

\begin{equation}
    M_q(\rho_X) \equiv \frac{1}{1-q}\log_2 \sum_{P \in \mathcal{P}_{|X|}} \Xi_P^q(\rho_X) \ - S_q(\rho_X) - \log_2 d,
\end{equation}
where $d \equiv 2^{|X|}$, $\mathcal{P}_{|X|}$ is the set of Paulis on $X$, and $\Xi_P(\rho_X) \equiv \frac{1}{d}\left[\Tr(P\rho_X)\right]^2$. The $q$th stabilizer R\'enyi entropy (SRE) is a magic monotone for $q \geq 2$~\cite{SRE_low_alpha, SRE_monotone}, introduced in~\cite{SRE}, that has been shown to be efficiently computable and measurable~\cite{SRE_algorithm, SRE_Markov, SRE_MPS} and studied in a wide range of physical systems~\cite{SRE_Ising, SRE_gravity, SRE_CFT, SRE_lattice_gauge, SRE_SYK, SRE_Rydberg}. We study the stabilizer R\'enyi-2 entropy

\begin{equation}\label{eqn:M2}
    M_2(\rho_X) = -\log_2\left(\Tr\left(Q \rho_X^{\otimes 4} \right) \right) - S_2(\rho_X) - \log_2 d,
\end{equation}
where $Q \equiv \frac{1}{d^2}\sum_{P \in \mathcal{P}_{|X|}} P^{\otimes 4}$.

The authors study the behavior of quantities averaged over Haar-random encodings $U$. To allow for tractable analytic results, they consider the  ``annealed" average. Let $\hat{\rho}_{U} \equiv \bra{0_{\overline{X}}} \mathcal{U}^\dagger \mathcal{E} \mathcal{U}(\rho_0)\ket{0_{\overline{X}}}$, the unnormalized, post-selected final state. Then $\rho_X = \frac{\hat{\rho}_U}{\Tr(\hat{\rho}_U)}$. For a given encoding $U$, each quantity of interest is a function $f(\frac{\hat{\rho}_U}{\Tr(\hat{\rho}_U)})$. The ``quenched," or exact, expectation value of each quantity is $\overline{f} \equiv \mathbb{E}_U f(\frac{\hat{\rho}_U}{\Tr(\hat{\rho}_U)})$, whereas the annealed average is $\tilde{f} \equiv f\left(\frac{\mathbb{E}_U \hat{\rho}_U}{\mathbb{E}_U\Tr(\hat{\rho}_U)}\right)$. In~\cite{enc_dec_original}, it is shown that the deviation between the two falls exponentially with system size, and close numerical agreement holds for system sizes as low as $N=8$.

\begin{figure*}[t]
    \centering
    \begin{subfigure}[t]{0.45\textwidth}
        \vspace{0pt}
        \centering
        \includegraphics[width=\textwidth]{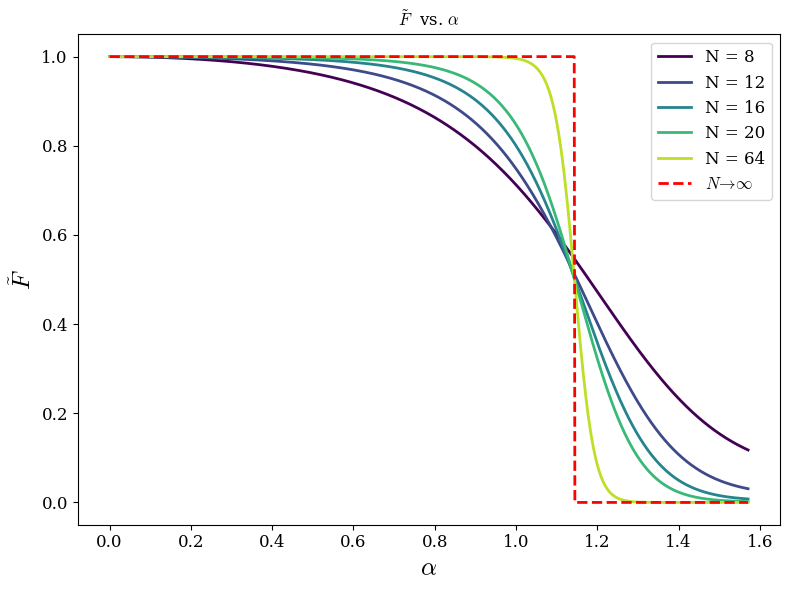}
        \caption{Fidelity versus $R_Z$ rotation angle $\alpha$ for various $N$ and the $N\rightarrow \infty$ limit.}
        \label{fig:fid_analytic}
    \end{subfigure}
    \hfill
    \begin{subfigure}[t]{0.45\textwidth}
        \vspace{0pt}
        \centering
        \includegraphics[width=\textwidth]{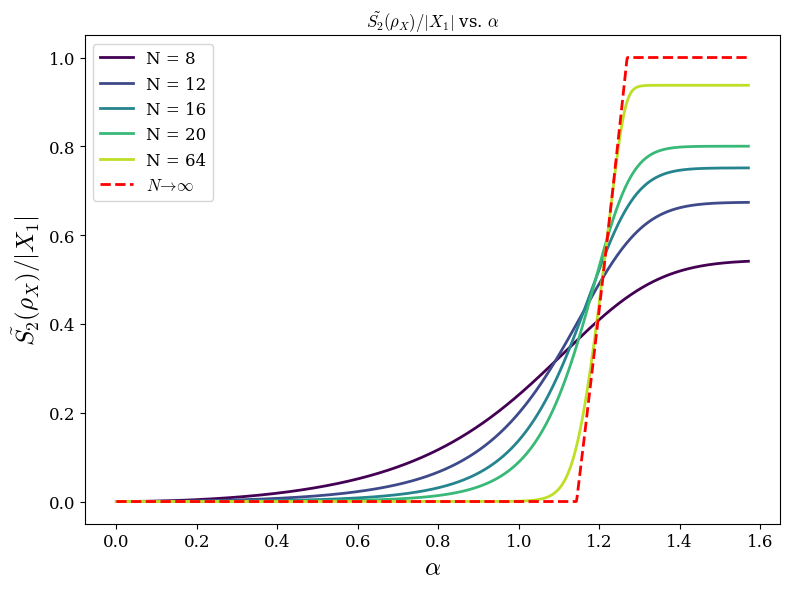}
        \caption{Entanglement entropy versus $R_Z$ rotation angle $\alpha$ for various $N$ and the $N\rightarrow \infty$ limit.}
        \label{fig:ee_analytic}
    \end{subfigure}
    \caption{Analytic curves for fidelity and entanglement entropy versus $R_Z$ rotation angle.}
    \label{fig:analytic_fid_and_ee}
\end{figure*}

We now state the analytic result for $\tilde{F}$:

\begin{equation}
    \tilde{F} = \frac{(2^N - 1)(2^N \cos^{2N}(\alpha/2) +1)}{2^N \cos^{2N}(\alpha/2)(2^N - 2^k) + 2^{N+k} -1},
\end{equation}
For brevity, we omit the lengthy analytic formulas for $\tilde{S_2}(\rho_X)$ and $\tilde{M_2}(\rho_X)$ at finite $N$. We plot finite-$N$ fidelities and entanglement entropies in \cref{fig:analytic_fid_and_ee}.

In the thermodynamic limit, $N \rightarrow \infty$, an ``error-resilience" phase transition occurs between an ``error-protecting phase" (EPP) in which the initial state $\ket{0_X}$ is recovered with certainty, and an ``error-vulnerable phase" (EVP) in which the state is entirely unrecoverable. The order parameter $\tilde{F}$ behaves as a step function from $1$ to $0$ between the phases, transitioning at $\alpha_c(r) = 2\arccos(2^{(r-1)/2})$, where $r \equiv k/N$ is the code ``rate." We take $r = 1/2$, for which $\alpha_c(1/2) \approx 1.14$.

The quantities $\tilde{S_2}(\rho_X)$ and $\tilde{M_2}(\rho_X)$ are also order parameters of the EPP-EVP transition, behaving in the thermodynamic limit as:

\begin{widetext}
\begin{equation}
\tilde{S_2}(\rho_X)/|X_1| = 
\begin{cases}
  0, & \alpha < \alpha_c(r) \\
  \frac{-4(\log_2[\cos(\alpha/2)] + (1-r)/2)}{r_1}, & \alpha_c(r) \leq \alpha < \alpha_X^{\text{ent}}(r), \\
  1, & \alpha \geq \alpha_X^{\text{ent}}(r),
\end{cases}
\end{equation}

and 

\begin{equation}
\tilde{M_2}(\rho_X)/N = 
\begin{cases}
  0, & \alpha < \alpha_c(r) \\
  8\log_2 \left[ \frac{\cos(\alpha_c(r)/2)}{\cos(\alpha/2)} \right], & \alpha_c(r) \leq \alpha <  2\arccos \left(2^{-r/8} \cos(\alpha_c(r)/2) \right), \\
  r, & \alpha \geq 2\arccos \left(2^{-r/8} \cos(\alpha_c(r)/2) \right),
\end{cases}
\end{equation}
\end{widetext}
where $X_1$ is the smaller of the subsystems of $X$, $r_1 = |X_1|/N$, and $\alpha_c^{\text{ent}}(r) = 2\arccos\left(2^{\left[ (2(r-r_1-1)+r_1)/4 \right]} \right)$. For $r = 1/2 \text{ and} \ r_1 = 1/4$, $\alpha_c^{\text{ent}}(1/2) \approx 1.27$. In the EVP, the final logical state becomes highly complex and nonlocal, transitioning from the product state $\ket{0_X}$ to one that is extensively entangled and classically difficult to simulate. Past $\alpha_c$, the coherent noise outcompetes random encoding and fully delocalizes the state.

The phase transition universality class is determined by Taylor expanding each order parameter, at finite $N$, about $\alpha_c$:

\begin{equation}\label{fidelity_scaling}
    \tilde{F} \approx \left(1 + \exp\left[\tan(\alpha_c/2)N(\alpha - \alpha_c) \right] \right)^{-1},
\end{equation}

\begin{equation}\label{ee_scaling}
    \tilde{S_2}(\rho_X) \approx 2\log_2\left(1 + \exp \left[\tan(\alpha_c/2)N(\alpha - \alpha_c) \right] \right),
\end{equation}

\begin{equation}\label{magic_scaling}
    \tilde{M_2}(\rho_X) \approx 4\log_2 \left[1 + \exp\left(\tan(\alpha_c/2)N(\alpha - \alpha_c) \right) \right].
\end{equation}
With the finite-size scaling ansatz $O = f(N^{1/\nu}(\alpha-\alpha_c))$ for order parameter $O$, the critical parameters are $\nu^{(1)} = \nu^{(2)} = \nu^{(3)} = 1$, where indices $1,2,3$ refer to fidelity, entanglement, and non-stabilizerness respectively.

\section{Experimental implementation of the theoretical model}\label{experiment_implement}

The fidelity and entanglement entropy phase transitions occur in any encoding-decoding family whose encoding unitaries $U$ are drawn from a unitary 2-design and 4-design respectively, while the non-stabilizerness transition occurs in 8-designs. One-dimensional brickwork circuits composed of Haar-random two-qubit gates asymptotically form $t$-designs in depth linear in system size $N$~\cite{t_design_lin_depth_1}. We therefore construct linear-depth brickwork circuits for experimental implementation. To maintain circuit depths meaningfully achievable on current devices, we set the brickwork layer depth \textit{exactly} equal to $N$.

Existing devices natively implement highly limited sets of two-qubit gates. To approximately construct the aforementioned brickwork circuits without reaching prohibitive depth, we generate circuits in which each layer is composed of random single-qubit gates followed by a fixed, device-native two-qubit gate on alternating qubit pairs. Tailoring this to the \texttt{ibm\_pittsburgh} Heron r3 device, whose native two-qubit gate is $CZ$~\cite{ibm_pittsburgh_compute_resource}, we implement random brickwork circuits of two-qubit-gate depth $N$, in which each layer is composed of a single-qubit layer of (native) $SX$ ($\sqrt{X}$) and $R_{Z}(\phi)$ with uniformly random $\phi$ on each qubits, followed by two-qubit $CZ$ gates. An arbitrary single-qubit unitary $U(\theta, \phi, \lambda)$ can be written as $R_Z(\phi)SX R_Z(\theta) SX R_Z(\lambda)$ \cite{virtual_rz}. Our single-qubit gates are $U(\phi) = R_{Z}(\phi)SX$, which we employ to reduce circuit depth. 

\begin{figure*}[t]
    \centering
    \begin{subfigure}[t]{0.45\textwidth}
        \vspace{0pt}
        \centering
        \includegraphics[width=\textwidth]{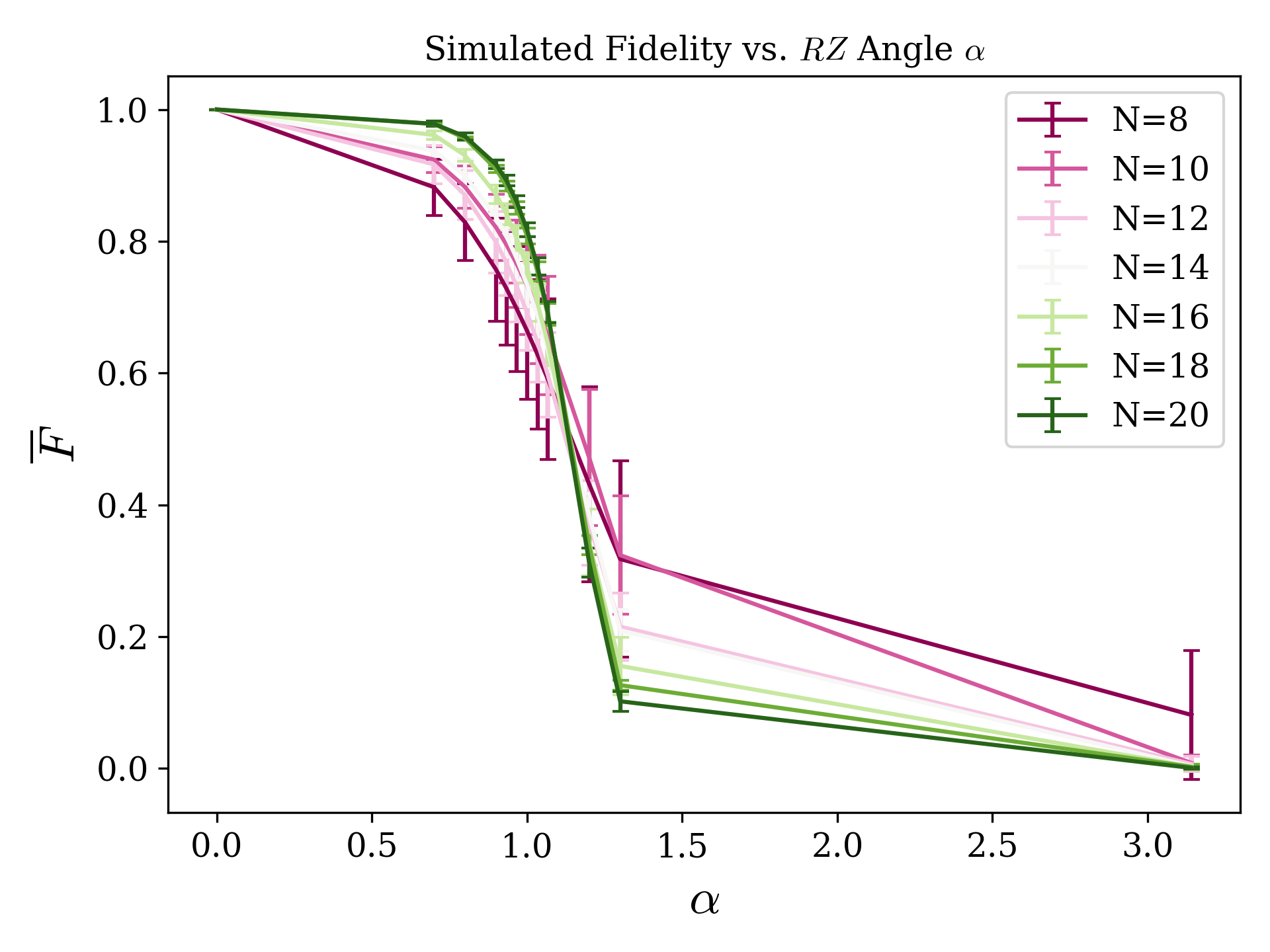}
        \caption{Fidelity versus $R_Z$ rotation angle $\alpha$.}
        \label{fig:fid_sim}
    \end{subfigure}
    \hfill
    \begin{subfigure}[t]{0.45\textwidth}
        \vspace{0pt}
        \centering
        \includegraphics[width=\textwidth]{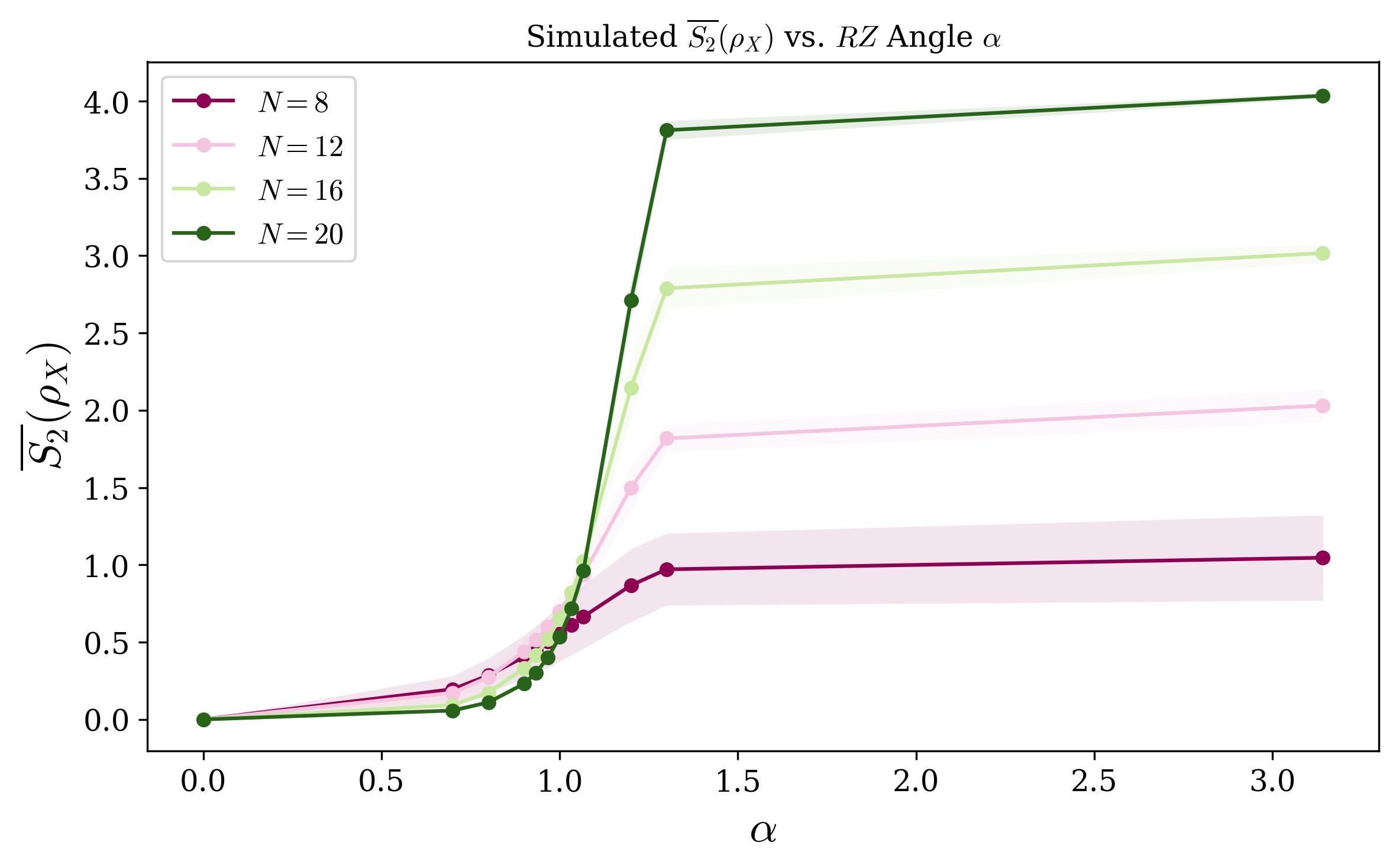}
        \caption{Entanglement entropy versus $R_Z$ rotation angle $\alpha$.}
        \label{fig:ee_sim}
    \end{subfigure}
    \caption{Fidelity and entanglement entropy versus $R_Z$ rotation angle for simulated, noiseless circuits. Each data point is the average of $4$ encoding circuits.}
    \label{fig:sim_fid_and_ee}
\end{figure*}

Our circuits do not implement Haar-random brickwork and the aforementioned linear-depth results hold asymptotically, with constant prefactors upper bounded by values far greater than 1. Depth-$N$ circuits may fail to yield results consistent with the Haar average. Before implementation on a real device, we tested the consistency of our construction with the Haar average via numerical simulations. For each $N$, using the above construction, we generated four circuits to implement on hardware. We use the same four encodings at each $\alpha$, which reduces the effect of inter-encoding variance leading to unexpected blips between $\alpha$ values and makes device implementation easier. We also classically simulated these circuits under various $\alpha$. Due to self-averaging, four circuits are enough to capture salient behavior of the average while remaining efficiently implementable on real devices. The simulated results are plotted in \cref{fig:sim_fid_and_ee}.

\begin{figure*}[t]
    \centering
    \includegraphics[width=0.6\textwidth]{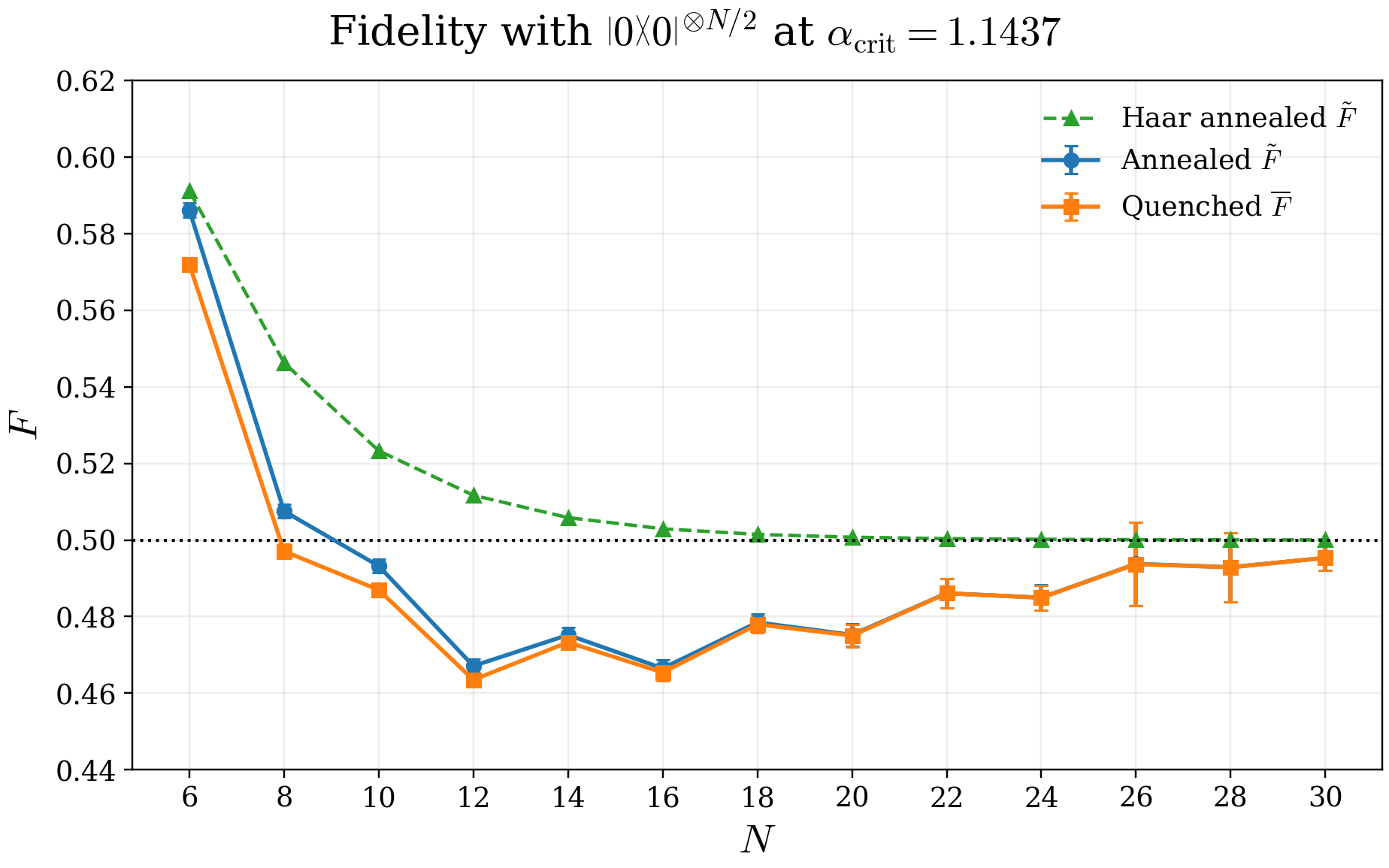} 
    \caption{Annealed and quenched fidelities of the experimental circuit ensemble at $\alpha = 1.1437$ plotted against $N$. The Haar-random ensemble annealed fidelity plot is displayed for comparison. As $N$ increases, the experimental ensemble fidelities approach $1/2$. Error bars reflect $\pm1$ standard deviation from mean. The number of simulated circuits per $N$ is displayed in \cref{tab:num_sim_fid_alphacrit}.}
    \label{fig:sim_crit_fidelity_vs_n}
\end{figure*}

To gain some insight into the large-$N$ behavior of this circuit ensemble, we observe the mean fidelity at the Haar-random $\alpha_c \approx 1.1437$. In the Haar case, $\tilde{F}(\alpha_c) \geq 1/2 \ \forall N$ and approaches $1/2$ as $N$ increases. We simulate circuits drawn from our experimental ensemble for $N$ from $6$ to $30$ and see, in \cref{fig:sim_crit_fidelity_vs_n}, that after an initial dip below $1/2$, $\tilde{F}(\alpha_c)$ and $\overline{F}(\alpha_c)$ approach $1/2$ as $N$ increases. This provides some evidence that the ensemble approaches the Haar-random results as $N \rightarrow \infty$. It is important to note, though, that since the ensemble is \textit{not} truly Haar-random for any finite $N$, we thereby expect the effective $\alpha_c$ modeled on finitely many $N$ to increase towards $1.1437$ as the dataset includes larger and larger $N$. This helps account for the discrepancy between the Haar-random ensemble $\alpha_c$ and the finite-$N$ effective $\alpha_c$ posteriors from both the simulated and experimental data.

\begin{table}[ht]
    \centering
    \caption{Number of simulated circuits per $N$ for the plot in \cref{fig:sim_crit_fidelity_vs_n}.}
    \label{tab:num_sim_fid_alphacrit}
    \begin{tabular}{cc}
        \toprule
        $N$ (system size) & \# Circuits \\
        \midrule
        6  & 30,000  \\
        8  & 20,000  \\
        10  & 10,000  \\
        12  & 5,000  \\
        14  & 2,500  \\
        16  & 1,000  \\
        18  & 500  \\
        20  & 150  \\
        22 & 40  \\
        24 & 20 \\
        26 & 5 \\
        28 & 5 \\
        30 & 3 \\
        \bottomrule
    \end{tabular}
\end{table}

\begin{figure*}[t]
    \centering
    \includegraphics[width=0.6\textwidth]{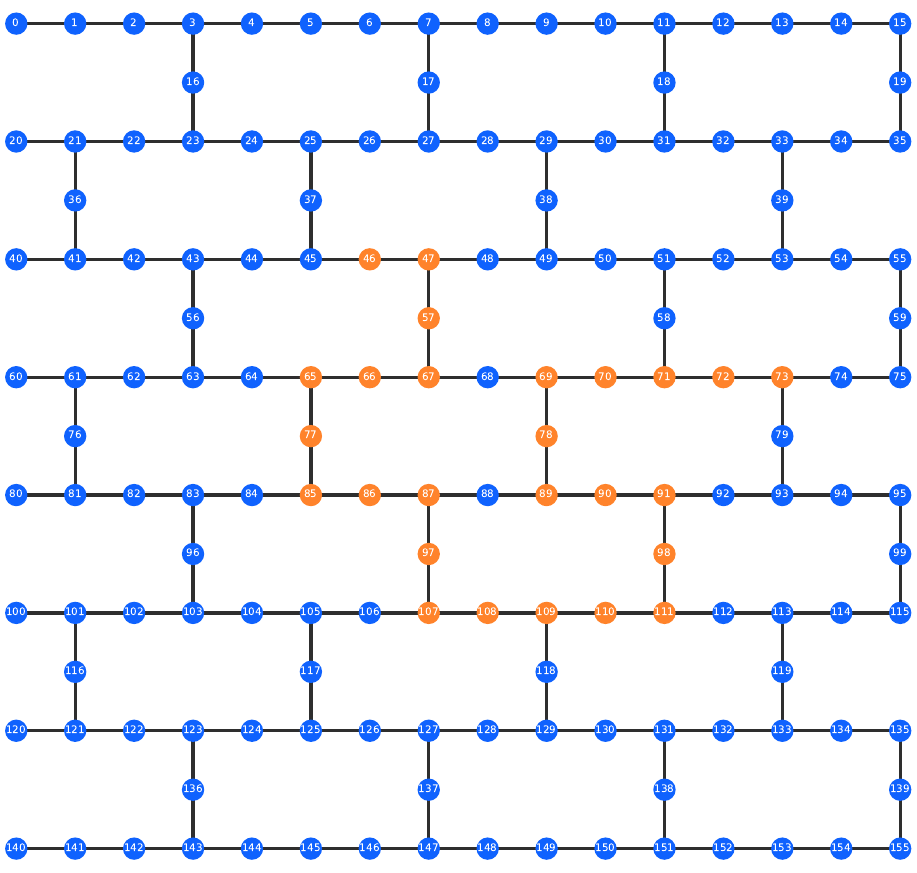} 
    \caption{Connectivity graph of the \texttt{ibm\_pittsburgh}
    processor, depicting the heavy-hex lattice structure employed by
    IBM Heron r3 devices.  The embedding for the $N = 26$ experiment is shown by orange qubits, while blue qubits were not used in the experiment.  Experiments conducted for system sizes $N = 8$ through $N = 24$ use a 1D chain subset of the orange qubits.}
    \label{fig:ibm_pittsburgh_connectivity}
\end{figure*}

As discussed, the coherent noise rotation axis does not affect the theoretical phase transition for Haar-random encodings, as only the rotation angle appears in any expressions. However, IBM devices apply virtual $R_{Z}(\phi)$ gates~\cite{virtual_rz}, implemented by adding phase offsets $\phi$ to the classical control frame, requiring no additional microwave pulse or increasing the circuit duration~\cite{virtual_rz}. We thus inject intentional coherent noise via $R_{Z}(\alpha)$ gates following the encoding unitary, since the error introduced by these gates is negligible. Technically, our non-Haar circuit ensemble's average behavior likely depends on the Bloch axis of the injected coherent noise. Since our experimental setup employs a single Bloch axis regardless, we use the native $Z$ axis. 

% Thus, the virtual $Z$ gate does not introduce the pulse-specific errors normally associated with a physically driven gate. Nevertheless, its effective implementation depends on maintaining a stable phase relationship between the qubit and the master clock or local oscillator that defines the rotating control frame. Phase fluctuations in this reference can cause the tracked frame to drift relative to the qubit, producing dephasing and phase errors that become observable when subsequent phase-sensitive operations are applied~\cite{ball2016masterclock}. In contrast, physically driven gates are additionally susceptible to errors arising during the applied pulse, including Stark-shift-induced rotation errors, population leakage into higher transmon levels, and decoherence accumulated over the finite gate duration~\cite{virtual_rz}. Master-clock phase noise is not unique to virtual $Z$ gates and can also degrade physically driven operations; the distinguishing advantage of the virtual implementation is the absence of an additional physical pulse and its associated error mechanisms. 

\section{Error mitigation}\label{error_mitigation}

\subsection{Zero-noise Extrapolation Implementation}
Zero-noise extrapolation is a method to reduce the effects of gate noise on experimental data.  Typically, some base circuit $U$ is implemented for multiple scaled values $\{\lambda_i\}_{i=1}^{n}$.  For each scaled value $\lambda_i$, the circuit length is extended by a factor of $\lambda_i$ while the overall unitary implemented by the circuit equals $U$.  For example, a circuit scaled by $\lambda = 3$ could be implemented as $U U^\dagger U$.  Without noise, this circuit implements $U$, but on a real device, extending the length of the circuit exacerbates the effects of the noise on the system.  Once data are collected for multiple $\lambda_i$-scaled circuits, a regression can estimate the measurement distribution of the noiseless ($\lambda = 0$) circuit.  

There are typically two methods used to extend the length of a base circuit: global folding and local folding.  In global folding, the entire circuit block, excluding measurements, is inverted and appended, to the base circuit to extend the length of the circuit.  Structurally, it looks as though the circuit $U$ is extended by having an integer number of blocks $U^{\dagger} U$ appended onto it.  Local folding takes each gate $g$ in $U$ and adds some number of sequences of $g^{\dagger}g$ after it to extend the length of $U$.  To integrate ZNE into our experiment, in which we only wanted to fold the encoding and decoding layers without altering the injected coherent noise, we wrote our own codebase. In our implementation, we opt for the global folding method to implement scaled circuits for $\lambda \in\{1, 3, 5\}$, where $\lambda = 1$ represents the base encoding circuit.  We then use quadratic Richardson extrapolation~\cite{Temme_2017} to interpolate a set of quadratic functions to the measured bit strings at each $\lambda$.  This set of quadratic functions is used to estimate the measurement distribution of each circuit in the zero-noise limit. As seen in \cref{fig:error_mitigation_comparison}, our implementation of ZNE is effective at mitigating the effects of circuit noise.

\subsection{M3 Measurement Error Mitigation}
We further utilize IBM's M3 measurement error mitigation package to reduce the impact of SPAM errors on our experimental data~\cite{Nation_2021}.  Our data collection pipeline is set up such that prior to each experimental run, we first perform an M3 calibration experiment on the qubits utilized by our experiment.  This serves to give us current device behavior at the time of data collection, such that the effects of the M3 package are maximized.  Experimental data is then collected immediately after the M3 calibration experiment to prevent drift or device recalibration from damaging the M3 measurement error mitigation technique.

\subsection{Efficacy of Error Mitigation}

\begin{figure*}[t]
    \centering
    \includegraphics[width=0.6\textwidth]{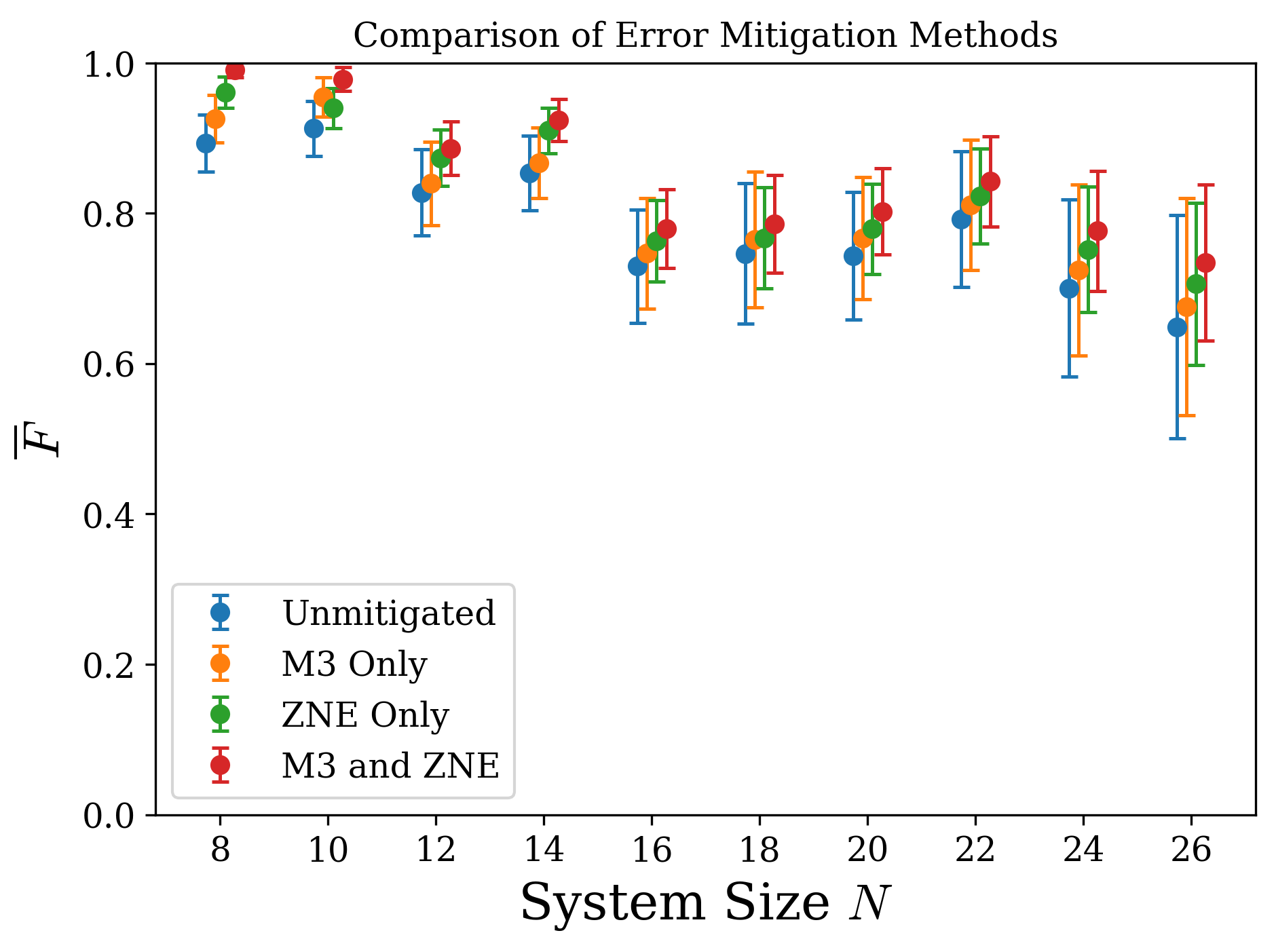} 
    \caption{We plot the experimental fidelity $\overline{F}$ of the system with respect to the system size $N$ after applying the circuit $UU^{\dagger} = \mathbb{I}$ ($\alpha = 0$) to the input state $|0_X \rangle$.  Each error mitigation strategy alone improves the fidelity compared to the unmitigated data, and combining both M3 and ZNE leads to the best overall performance.}
    \label{fig:error_mitigation_comparison}
\end{figure*}

A simple benchmark of ZNE and M3 efficacy in our experiments is the fidelity with $\ket{0_X}$ at noise angle $\alpha = 0$. In a noiseless system, every encoding-decoding circuit implements $U\mathbb{I}U^\dag = \mathbb{I}$, yielding $F = 1$. Gate and SPAM noise generally reduce this value. We look at $\alpha = 0$ fidelities at each system size $N$ for circuits with no error mitigation, with M3, with ZNE, and with both applied. We see in \cref{fig:error_mitigation_comparison} that the highest fidelities are achieved using both methods.  We considered dynamical decoupling (DD) to reduce correlated noise during idling, but our experiments revealed that it did not improve circuit performance.  There are two primary reasons for this: first, the tunable couplers on the IBM Heron devices significantly reduce crosstalk, and second, the brickwork structure in our circuits means that only boundary qubits having very short idle periods.

\section{Measuring the Order Parameters}\label{measurements}

The logical state fidelity with $\ket{0_X}$ is simply the fraction of post-selected measurement outcomes which give the bit string $\ket{0_X}$. In our experiments, we execute $30{,}000$ shots per circuit for fidelity measurement. The entanglement entropy and non-stabilizerness are functions of $\rho_{|X_1|}^2$ and $\rho_X^4$, respectively. In principle, one could measure these quantities with multi-copy observables of $\rho_{|X_1|}^{\otimes 2}$ and $\rho_X^{\otimes 4}$. However, such measurements require numerous swaps and permutations between qubits on different copies of the state and are therefore prohibitively complex for our qubit counts on NISQ (noisy intermediate-scale quantum) era superconducting devices. To quantify the entanglement entropy and non-stabilizerness, we instead employ single-copy, randomized measurement (RM) protocols.

We note that $S_2(\rho_{|X_1|})$ is only a true entanglement entropy when $\rho_X$ is a pure state. Furthermore, the expression for $M_2(\rho_X)$ in \cref{eqn:M2} involves subtracting $S_2(\rho_X)$, a quantity we have not experimentally observed. On the noisy devices used in our experiments, $\rho_X$ is a mixed state. Referring to $S_2(\rho_{|X_1|})$ as the R\'enyi-2 entanglement entropy of $\rho_X$ is technically incorrect, though we continue to do so for clarity and consistency with \cref{theory}; we also set $S_2(\rho_X) = 0$ in computing $M_2(\rho_X)$ with experimental data. We elaborate on how we partially correct for the impurity of $\rho_X$ in \cref{data_cleaning}.

\subsection{Measuring Entanglement Entropy}

\subsubsection{Protocol}

Let $u = \bigotimes\limits_{i=1}^{n} u_i$ be a random product unitary on $n$ qubits, where $\forall i, \ u_i$ is independently drawn from the Haar measure. Also let $s \in \{0, 1 \}^n$ denote a bit string, and let $P_u(s) \equiv \bra{s}u\rho u^\dagger \ket{s}$. Then, for an $n$-qubit state $\rho$, 

\begin{equation}
\mathbb{E}_u X_\rho(u) = \Tr(\rho^2) \equiv \gamma(\rho),
\end{equation}
where
\begin{equation}\label{RM_purity}
    X_\rho(u) \equiv 2^n \sum\limits_{s, s'} (-2)^{-D\left[s,s' \right]} P_u(s) P_u(s').
\end{equation}
$D$ denotes the Hamming distance. We set $n \equiv |X_1| = N/4$ and apply RMs over $n$ qubits, take the sample mean over these to yield an estimator of $\gamma(\rho_{|X_1|})$ (represented as $\hat{\gamma}(\rho)$), then take $\hat{S_2}(\rho_X) \equiv -\log_2(\hat{\gamma}(\rho_{|X_1|}))$. We then, for an $\left(N, \alpha \right)$ pair, take our entanglement entropy estimator to be $\frac{1}{N_U} \sum_i \hat{S_2}^{(i)}(\rho_X)$, averaging over encodings $U$. Via Jensen's inequality, this introduces a bias in $\hat{S_2}(\rho_X)$, similar to the annealed average in \cref{theory}, which we address in \cref{avg_comparison}.

Up to global phase, any single-qubit unitary can be realized by the parameterization $u_i = R_Z(\gamma_i)R_Y(\theta_i)R_Z(\phi_i)$~\cite{IS, RM_probing_entanglement}. The Haar measure on these coordinates is $du_i = \frac{1}{8 \pi^2}\sin\theta_i d\theta_i d\gamma_i d\phi_i$, with $\theta_i \in \left[0, \pi\right]$ and $\gamma_i, \phi_i \in \left[0, 2\pi\right)$. Since all measurements are performed in the computational ($Z$) basis, the final $R_Z(\gamma_i)$ does not affect observed outcomes, and we may omit it. Sampling $\theta_i$ with $p(\theta_i) = \frac{1}{2}\sin\theta_i$ on $\left[0, \pi \right]$ and $\phi_i$ uniformly on $\left[0, 2\pi \right]$ then inserting into the above parameterization gives Haar-random $u_i$.

\subsubsection{Approximate state ansatz}

Incorporating prior knowledge of a quantum state can reduce the variance of the randomized measurement purity estimator. Two methods, the “common randomized measurements” approach (CRM)~\cite{CRM} and importance sampling (IS)~\cite{IS}, rely on possessing a classical description of a state $\sigma$ that approximates the true state $\rho$. For each method, the nearer $\sigma$ is to $\rho$, the lower the variance of the corresponding purity estimator. We describe each method below.

To construct our approximate state, we rely on the fact that we implement the randomized measurements following the fidelity experiments. These experiments give full computational basis empirical distributions for each circuit. However, we incorporate only the fidelity of the final logical state with $\ket{0_X}$ into our approximate state for several reasons. Consider an approximate state $\sigma$ constructed by matching the diagonal elements of $\rho$ in the computational basis. First, the computational basis distribution does not directly constrain the trace distance between the states or the relationship between their purities; rather, it merely implies that computational basis measurements cannot distinguish between $\sigma$ and $\rho$. The same cannot be said of other bases. Second, the theoretical result describes the fidelity between the initial and final state, without addressing specific elements of the orthogonal complement of the initial state. The fidelity is therefore a salient feature of the final state, unlike the computational basis probabilities. Finally, we want a simple ansatz that is computationally tractable for large system sizes, and constructing one from exponentially many bit strings would vastly increase its complexity.

With these considerations in mind, we propose an ansatz of the form $\sigma(p) = (1-p)\ket{0_{X_1}}\bra{0_{X_1}} + p\frac{\mathbb{I}}{2^{|X_1|}}$. For both variance reduction methods, we consider $p$ values ranging from 0 to 1 and numerically observe which ones provide the largest variance reduction. We then compare these to the observed fidelity of the state with the all-0 state.

\subsubsection{Common Randomized Measurements}

Inspired by~\cite{CRM}, we consider a CRM setup specifically for purity estimation. Given an approximate state $\sigma$, we can compute both $\Tr(\sigma^2)$ and $X_\sigma(u)$ for any RM $u$. CRM relies on the observation:
\begin{widetext}
\begin{equation}
\mathbb{E}_u \left[ X_\rho(u) - a(X_\sigma(u) - \Tr(\sigma^2)) \right] = \mathbb{E}_u \left[ X_\rho(u) - a\mathbb{E}_u X_\sigma(u)\right] + a\Tr(\sigma^2) = \Tr(\rho^2) - a\Tr(\sigma^2) + a\Tr(\sigma^2) = \Tr(\rho^2).
\end{equation}
\end{widetext}
The estimator $X_\rho(u) - a(X_\sigma(u) - \Tr(\sigma^2))$ is unbiased and has a variance of $\Var(X_\rho) + a^2 \Var(X_\sigma) - 2a\Cov(X_\rho, X_\sigma)$. It is simple to see that the variance-minimizing value of $a$ is $a^* = \Cov(X_\rho, X_\sigma)/\Var(X_\sigma)$, but an $X_\sigma$ highly correlated with $X_\rho$ generally reduces variance for $a \in (0, 1]$. For our experiments, we consider only $a = 1$ for two reasons. First, we expect to identify highly effective ans\"atze, namely ones for which $X_\sigma$ is very close to $X_\rho$, which implies that $\Cov(X_\rho, X_\sigma) \approx \Var(X_\sigma)$. Second, since we do not have access to a classical description of $\rho$, we must compute the empirical $\Cov(X_\rho, X_\sigma)$ via the same randomized measurements we use to estimate $\Tr(\rho^2)$. This introduces bias in the CRM estimator and the variance of the estimator $\hat{a^*}$ can itself increase the CRM estimator's variance. 

This method has the benefit of being applicable entirely in post-processing. One can conduct the randomized measurements all at once and then retroactively decide which approximate state gives the greatest variance reduction. The division of workflows makes this method especially simple and efficient to implement.

\subsubsection{Importance Sampling}

The function $X_\sigma$ is positive, taking values in the range $\left[2^{-n}, 2^n \right]$ for an $n$-qubit state. Given that $\mathbb{E}_u X_\sigma(u) = \int_u X_\sigma(u)\, du = \Tr(\sigma^2)$, the normalized function $\frac{X_\sigma(u)}{\int_u X_\sigma(u)\, du} = \frac{X_\sigma(u)}{\Tr(\sigma^2)}$ is a probability density function over $u$, which we refer to as $p_{IS}(u)$.

Straightforwardly, 
\begin{equation}
\int_u \left( \frac{X_\rho(u)}{p_{IS}(u)} \right) p_{IS}(u)\, du = \int_u X_\rho(u)\, du = \Tr(\rho^2).
\end{equation}

Over the distribution $p_{IS}(u)$, the function $\frac{X_\rho(u)}{p_{IS}(u)}$ has the variance 
\begin{widetext}
\begin{equation}
\mathbb{E}_{p_{IS}(u)} \left( \frac{X_\rho(u)}{p_{IS}(u)} \right)^2 - \left(\mathbb{E}_{p_{IS}(u)}\frac{X_\rho(u)}{p_{IS}(u)}\right)^2 = \int_u \frac{X_\rho^2(u)}{p_{IS}(u)}\, du - \Tr(\rho^2)^2.
\end{equation}
\end{widetext}

Note that in the case of perfect importance sampling, i.e., \ $p_{IS}(u) = \frac{X_\rho(u)}{\int_u X_\rho(u)\, du}$, the variance is $\int_u \frac{X_\rho^2(u)}{ \frac{X_\rho(u)}{\int_u X_\rho(u)\, du}}\, du - \Tr(\rho^2)^2 = 0$. 

Using Monte Carlo methods~\cite{monte_carlo_metropolis_original, monte_carlo_textbook}, given an approximate state $\sigma$, one can sample from the distribution $p_{IS}(u)$. The difficulty of performing such sampling depends on the complexity of $X_\sigma(u)$. For example, in~\cite{IS}, the authors demonstrate the efficacy of learning a simplified neural network approximation of $p_{IS}(u)$, and then sampling from it via the Metropolis algorithm. By sampling from $p_{IS}(u)$ then recording the values of $p_{IS}(u)$ and empirically observing $X_\rho(u)$, one can construct the purity estimator $\frac{1}{N_u}\sum_u \frac{ \hat{X}_\rho(u) }{p_{IS}(u)}$. It was numerically observed in~\cite{IS} that for certain classes of true and approximate states, this method can lead to an exponential reduction in measurement overhead, compared to uniform sampling, required to achieve a fixed variance.

Our aforementioned ansatz for $\sigma$ yields an $X_\sigma$ that is efficient to sample from directly, which we now derive. In the following, we represent vectors as kets with bolded letters and use bra-ket notation for linear algebra. Consider the form of $X_\sigma(u)$ in \cref{RM_purity}. Let $\mathbf{v}$ and $\mathbf{w}$ be the computational basis probability vectors for the states $\ket{0}^{\otimes N/4}$ and $\left(\frac{\mathbb{I}}{2}\right)^{\otimes N/4}$, respectively. Let $F$ be the matrix defined by $F_{s, s'} = (-2)^{-D \left[s, s' \right]}$, and $f$ refer to the same matrix for a single qubit, so that $F = f^{\otimes N/4}$. By linearity and reality of the probability vectors, 

\begin{widetext}
\begin{equation}
\begin{split}
X_{\sigma(p)}(u) &= 2^{N/4} \left( \bra{(1-p)\mathbf{v} + p\mathbf{w}} F \ket{(1-p)\mathbf{v} + p\mathbf{w}} \right) \\
&= 2^{N/4} \left( (1-p)^2 \bra{\mathbf{v}} F \ket{\mathbf{v}} + p^2 \bra{\mathbf{w}} F \ket{\mathbf{w}} + 2p(1-p)\bra{\mathbf{v}} F \ket{\mathbf{w}} \right).
\end{split}
\end{equation}
\end{widetext}

Since $\frac{\mathbb{I}}{2^{N/4}} = \left(\frac{\mathbb{I}}{2}\right)^{\otimes N/4}$, we immediately see $\bra{\mathbf{w}} F \ket{\mathbf{w}} = \left( \bra{1/2, 1/2} f \ket{1/2, 1/2} \right)^{N/4} = \frac{1}{4^{N/4}}.$ Again using $f\ket{1/2, 1/2} = (1/4, 1/4),$ we see that $2p(1-p)\bra{\mathbf{v}} F \ket{\mathbf{w}} = 2p(1-p) \prod_i \bra{\mathbf{v}_i} f \ket{\mathbf{w}_i} = \frac{2p(1-p)}{4^{N/4}} \prod_i (v_i^0 + v_i^1) = \frac{2p(1-p)}{4^{N/4}}$ by normalization.

Finally, $\bra{\mathbf{v}} F \ket{\mathbf{v}} = \prod_i \bra{v_i} f \ket{v_i} = \prod_i \left((v_i^0)^2 + (v_i^1)^2 - v_i^0 v_i^1 \right)$.

Recall the random single-qubit measurement $R_Y(\theta_i)R_Z(\phi_i)$. Now, since the state $\sigma(p) = (1-p)\ket{0}\bra{0}^{\otimes N/4} + p\frac{\mathbb{I}}{2^{N/4}}$ is diagonal in the $Z$ basis, probabilities are invariant under application of any $R_Z$ gate. The purity estimator $X_\sigma(u)$ depends only on the $R_Y$ angles. We rewrite $R_Y(\theta_i) = e^{-i\frac{\theta_i}{2} Y}$ as $\cos\!\left(\frac{\theta_i}{2}\right)\mathbb{I} - i\sin\!\left(\frac{\theta_i}{2}\right)Y.$ Now, we apply this unitary to the state $\ket{0}\bra{0}^{\otimes N/4}$ to easily see the rotated-state bit probabilities $v_i^0(u) = \cos^2\!\left(\frac{\theta_i}{2}\right)$ and $v_i^1(u) = \sin^2\!\left(\frac{\theta_i}{2}\right)$.

Collecting terms, we get:

\begin{widetext}
\begin{equation}
X_{\sigma(p)}(u) = 2^{N/4} (1-p)^2 \prod_{i=1}^{N/4} \left( 1 - \frac{3}{4}\sin^2\!\left(\theta_i\right) \right) + \frac{1-(1-p)^2}{2^{N/4}}.
\end{equation}
\end{widetext}

This is a simple analytic function of the $R_Y$ rotation angles $\{\theta_i\}_{i=1}^{N/4}$. We divide $X_{\sigma(p)}(u)$ by $\Tr(\sigma(p)^2)$ to yield a probability distribution over $u$. We then convert it to a distribution over $\{\theta_i\}_{i=1}^{N/4}$ by multiplying by the above Jacobian $\prod_i \frac{1}{2} \sin\theta_i$. Since this function is simple and directly differentiable, we can easily feed it into the No-U-Turn Sampler (NUTS) algorithm~\cite{NUTS} to approximately draw samples from this distribution. NUTS is an extremely efficient extension of the Hamiltonian Monte Carlo algorithm that allows for rapid exploration of high-density regions of the target distribution. Because the true state is not diagonal in the $Z$ basis, we also apply an $R_Z$ gate on each qubit with uniformly randomly sampled angles $\{\phi_i\}_{i=1}^{N/4}$.

\subsubsection{Numerical comparison of CRM vs. IS}

Both CRM and IS have been demonstrated to significantly reduce the variance of purity estimation. For our system, in which we lack precise knowledge of the true logical state of any particular circuit, direct analytical comparisons between the methods are not possible. Before deciding which to use, we numerically compare them on noiseless simulations of our random encoding-decoding circuits, implemented in Qiskit.

For each N, we generate 10 random encodings which we use for each $R_{Z}(\alpha)$ value. We apply $R_{Z}(\alpha)$ angles $\alpha$ in $\{0, 0.7, 0.8, 0.9, 0.933, 0.967, 1, 1.033, 1.067, 1.2, 1.3, \pi \}$, concentrated in the transition region. We then plug in 100 values of $p$ evenly spaced between $0$ and $1$ (inclusive) for $\sigma(p)$. For each, we calculate the variance of the CRM and IS estimators, using the same number of measurements and shots per measurement as in our experiments (see \cref{ee_meas_counts}) and scaling shot counts by the experimental mean post-selection frequencies for each $(N, \alpha)$ pair in \cref{appendix_postselection}. We emphasize that we do not actually \textit{perform} importance sampling for these calculations, as doing so is time-intensive. Rather, we compute the “weights” $p_IS(u)$ for \textit{uniformly} sampled $u$. Note that the variance of our importance sampling purity estimator function $\frac{X}{p_{IS}}$ is:

\begin{widetext}
\begin{equation}
\Var_{p_{IS}} \frac{X}{p_{IS}} = \mathbb{E}_{p_{IS}} \frac{X^2}{p^2_{IS}} - \left( \mathbb{E}_{p_{IS}}  \frac{X}{p_{IS}} \right)^2 = \mathbb{E}_{uni} \frac{X^2}{p^2_{IS}} p_{IS} - \left( \mathbb{E}_{uni}  \frac{X}{p_{IS}} p_{IS} \right)^2 = \mathbb{E}_{uni} \frac{X^2}{p_{IS}} - \left( \mathbb{E}_{uni}  X \right)^2.
\end{equation}
\end{widetext}
With the weights $p_{IS}$ and our uniformly sampled $u$, we compute the expression in the last line as our estimator of the IS variance \footnote{We note this is a higher-variance estimator of the IS variance than the one obtained by actually performing IS and computing the observed variance.}.

\begin{figure}[h]
    \centering
    \includegraphics[width=\columnwidth]{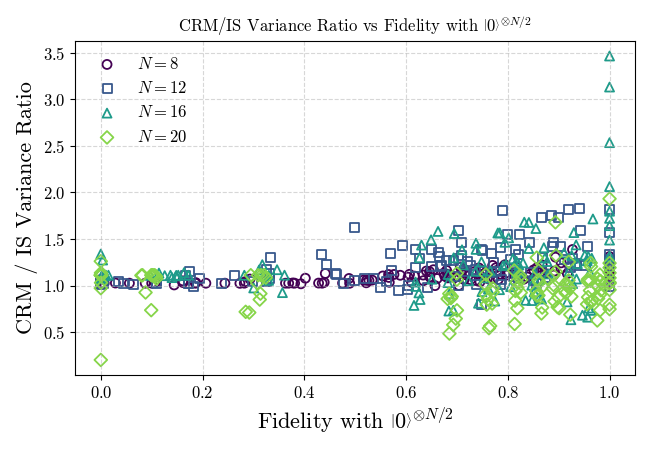}
    \caption{Ratio of CRM vs. IS purity estimator variances plotted against the fidelity of the underlying logical state with $\ket{0_X}$. For each $N$, $10$ encodings were simulated.}
    \label{fig:crm_is_ratio}
\end{figure}

For each $\alpha$, we record the lowest-variance $p$ value and corresponding variance for both CRM and IS, then plot the ratio of these variances against the true state fidelity in \cref{fig:crm_is_ratio}. We observe that for each fidelity, the ratio is typically slightly larger than or near $1$ for $N \in \{8, 12, 16\}$, meaning that the IS estimators have slightly lower variances. For $N = 20$, there is a wide range of ratios, and the advantage of IS over CRM is marginal to non-existent. However, to keep a streamlined workflow, we use IS for our device experiments for all $N$ values.

\subsubsection{Selecting the best IS approximate state}

\begin{figure}[h]
    \centering
    \includegraphics[width=\columnwidth]{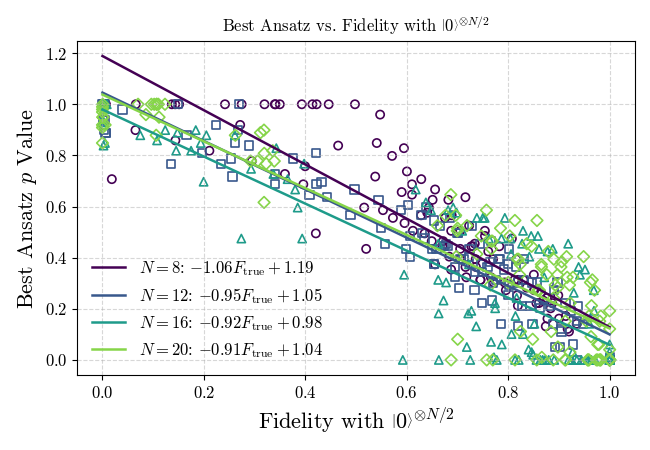}
    \caption{Lowest-variance IS ansatz $p$ value plotted against the fidelity of the underlying logical state with $\ket{0_X}$.}
    \label{fig:best_ansatz_p_is}
\end{figure}

For each $N$ and $R_{Z}(\alpha)$ value, we compute the average $\overline{F}$. We also compute the average best $p$ value for that $(N, \alpha)$ pairing. We then have pairs of $\left(\overline{F},  p_{\text{best}} \right)$, on which we perform linear regression separately for each $N$. The results are plotted in \cref{fig:best_ansatz_p_is}.

To account for some $y$-intercepts slightly exceeding $1$, and to keep the experimental workflow simple, we use $\sigma(p)$ with $p = 1-0.9\overline{F}$ as our approximate states for all $N$. These then get fed into the NUTS algorithm.

\subsubsection{Comparing IS to Uniform Measurements}

\begin{figure}[h]
    \centering
    \includegraphics[width=\columnwidth]{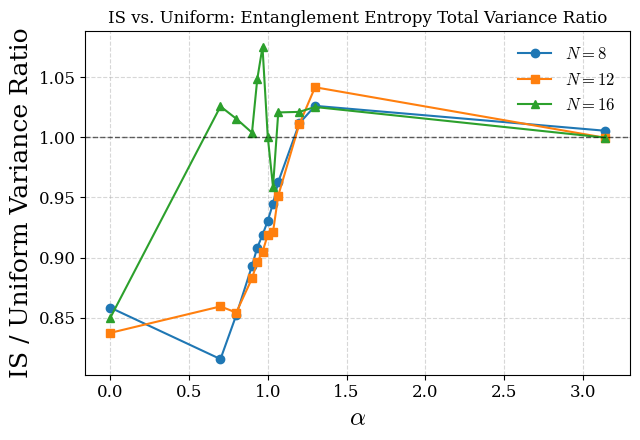}
    \caption{IS vs. uniform measurement variance ratio of the purity estimator over different simulated encodings. Shot counts approximately match the total post-selected shot counts in experiment.}
    \label{fig:is_vs_uniform_var_ratio}
\end{figure}

We now compare IS using an approximate state with $\sigma(1-0.9F)$ to selecting single-qubit measurements Haar randomly. For system sizes $N \in \{8, 12, 16\}$, we randomly select 500, 500, and 50 encodings, respectively, and exactly simulate the final logical state over the 12 aforementioned $\alpha$ values. We then sample measurements Haar-randomly, and sample shots from each measurement, for each $N$ according to \cref{ee_meas_counts} and the post-selection frequencies in \cref{appendix_postselection}. These results give us the Haar ``uniform" RM estimators for R\'enyi-2 entropy. With the same set of measurements, we compute the simulated IS estimators. For each $(N, \alpha)$ pair, we plot the IS versus uniform measurement variance ratio over the set of random encodings in \cref{fig:is_vs_uniform_var_ratio}. IS reduces the intra-encoding variance of the purity estimator due to finite shot counts. However, much of the total variance of the encoding-averaged means is due to inter-encoding variance (e.g., see \cref{fig:histogram_plots_normal}), which is the same for either measurement protocol. Therefore, the total variance ratio of IS to uniform measurements is not much less than 1.

These data indicate that at shot counts matching those in our experiment, IS may provide a small improvement over uniform sampling for smaller values of $\alpha$. This is in accordance with the facts that 1) IS has greater variance reduction as the number of shots increases and 2) the post-selection frequency is higher at smaller $\alpha$. We note that the values above 1 in \cref{fig:is_vs_uniform_var_ratio} may be spurious, due to the variance of the variance estimator being higher for simulated IS than actual IS (see \cref{fn:N18}); we likely do not have enough post-selected shots to consistently see meaningful signal in this ratio.

\subsection{Measuring Non-stabilizerness}

In~\cite{enc_dec_magic}, it was shown that this model undergoes a magic phase transition, in addition to the fidelity and entanglement entropy transitions, at the same $\alpha_c$, as quantified by the second and third stabilizer R\'enyi entropies (SREs). We restate that the second SRE of an $n$-qubit state can be written as:
\begin{equation}
M_2(\rho) = -\log_2 \Tr(Q \rho^{\otimes 4}) - S_2(\rho) - \log_2 d,
\end{equation}
where $Q \equiv \frac{1}{d^2} \sum_{P \in \mathcal{P}_n} P^{\otimes 4}$ and $d \equiv 2^n$.

% The $q$th SRE is a magic monotone for $q \geq 2$~\cite{SRE_low_alpha, SRE_monotone}, introduced in~\cite{SRE}, that has been shown to be efficiently computable and measurable~\cite{SRE_algorithm, SRE_Markov, SRE_MPS} and studied in a wide range of physical systems~\cite{SRE_Ising, SRE_gravity, SRE_CFT, SRE_lattice_gauge, SRE_SYK, SRE_Rydberg}. We state:

% \begin{definition}[Stabilizer R\'enyi Entropy]
% The $q$th Stabilizer R\'enyi Entropy of an $n$-qubit state $\rho$ is defined as:
% \begin{equation}
% M_q(\rho) = \frac{1}{1-q} \log_2 \sum_{P \in \mathcal{P}_n} \frac{\Tr(\rho P)^{2q}}{2^n},
% \end{equation}
% where $\mathcal{P}_n$ is the $n$-qubit Pauli group.
% \end{definition}

In~\cite{SRE}, it was proven that there exists a randomized measurement estimator of $M_2$ involving random product Clifford measurements. Let $c = \bigotimes_{i=1}^{n} c_i.$ Then

\begin{widetext}
\begin{equation}\label{SRE_RM_estimator}
-\log_2 \Tr(Q \rho^{\otimes 4}) = -\log_2 \mathbb{E}_c \left[   \sum_{s_1, s_2, s_3, s_4} \left( -2 \right)^{-\|s_1 \oplus s_2 \oplus s_3 \oplus s_4 \|} P_c(s_1)P_c(s_2)P_c(s_3)P_c(s_4) \right],
\end{equation}
\end{widetext}
where $\|\cdot\|$ denotes Hamming weight, $P_c(s) \equiv \bra{s}c\rho c^\dagger \ket{s}$, and each $c_i$ is drawn uniformly from the single-qubit Clifford group. We define $W_c(\rho)$ as the bracketed expression inside the expectation.
% \begin{equation}
% \begin{split}
% -\log_2 \Tr(Q \rho^{\otimes 4}) &= \mathbb{E}_c \Bigg[ -\log_2 \sum_{s_1, s_2, s_3, s_4}
%   (-2)^{-(s_1 \oplus s_2 \oplus s_3 \oplus s_4)} \\
%   &\qquad\qquad \times\, P_c(s_1)\, P_c(s_2)\, P_c(s_3)\, P_c(s_4) \Bigg].
% \end{split}
% \end{equation}

The single-qubit Clifford group has 24 elements. Every element acts on the Pauli basis by permuting the non-identity Paulis and flipping up to two of their signs~\cite{single_qubit_cliffords}. Flipping the sign of a Pauli operator just reassigns which eigenstates are labeled $+1$ and $-1$. The RM estimator in \cref{SRE_RM_estimator} depends only on the Hamming weight of a binary sum of four bit strings, and the binary sum of an even number of bit flips is $0$. This leaves at most 6 distinct single-qubit Cliffords (the Pauli permutations) for this protocol. Furthermore, following the application of the random product Clifford, the device performs a Z measurement on each qubit. This means that permutations can only affect the protocol insofar as they map Z to different Paulis, leaving just 3 distinct choices of single-qubit Cliffords: namely, we select the class representatives $\mathbb{I}$, $H$, and $HS^\dagger$, mapping Z to Z, X, and Y respectively. We uniformly sample from this set for the hardware experiment.

\subsection{U-statistics}

The above RM estimator formulas for purity and non-stabilizerness assume access to the true probabilities $P_u(s)$ for RM $u$ and bit string $s$. In real experiments, however, we can only access the \textit{observed} frequencies $\hat{P_u}(s)$. Let $N_M$ be the number of measurement shots for an RM $u$, and let $m$ and $m'$ refer to two particular shot indices. Both the purity and non-stabilizerness estimators are expectations of symmetric ``kernel" functions over 2-tuples and 4-tuples of bit strings, respectively. The minimum-variance unbiased estimator of each quantity, given $N_M$ shots, is obtained by averaging this kernel over all distinct shot tuples, which is precisely the U-statistic. 

For instance, the purity ``kernel" $2^n(-2)^{-D\left[s_m, \ s_{m'} \right]}$ is an unbiased estimator of $X_\rho(u)$ for an $n$-qubit $\rho$. The U-statistic of $X_\rho(u)$ given $N_M$ shots is:

\begin{equation}
    \hat{X_\rho}(u) = 2^n\binom{N_M}{2}^{-1} \sum_{\substack{m,\, m' = 1 \\ m > m'}}^{N_M} (-2)^{-D\left[s_m, \ s_{m'} \right]}.
\end{equation}

The non-stabilizerness U-statistic is:

\begin{widetext}
\begin{equation}
    \hat{W_c}(\rho) = \binom{N_M}{4}^{-1} \sum_{\substack{m_1,m_2,m_3,m_4 = 1 \\ m_1{>}m_2{>}m_3{>}m_4}}^{N_M} (-2)^{-\|s_{m_1} \oplus s_{m_2} \oplus s_{m_3} \oplus s_{m_4} \|}.
\end{equation}
\end{widetext}
The entanglement entropy and stabilizer R\'enyi entropy values throughout this manuscript are computed via the corresponding U-statistics.

\subsection{Comparison Between Averages}\label{avg_comparison}

The specific analytic formulas in \cref{theory} hold for annealed averages. However, our experimental averages are not of this form. Our fidelity averages are simple averages of the form $\frac{1}{N_U}\sum\limits_{i=1}^{N_U} F^{(i)}$, where $F^{(i)}$ is the observed fidelity given the encoding $U_i$. This is an unbiased estimator of $\mathbb{E}_U F\left(\frac{\hat{\rho_U}}{\Tr(\hat{\rho_U})} \right) = \overline{F}$, the quenched average. Our estimators of $S_2$ and $M_2$ correspond to neither the quenched nor annealed averages. Since we are estimating quantities of the \textit{post-selected} state and then taking the logarithm, we are introducing bias relative to the quenched average via Jensen's inequality. Since we are not separately averaging the numerator and denominator \textit{inside} the logarithm, we are not estimating the annealed average either. As discussed in~\cite{enc_dec_original}, however, the discrepancies between our averages and the annealed averages are exponentially small, with extremely close agreement even at present experimental system sizes. We keep the tilde notation for our entanglement entropy and non-stabilizerness estimators.

\subsection{Number of Randomized Measurements}

Due to constraints of the quantum device platform, it was most efficient to submit circuits for a single $N$ at a time. To then maximize the number of shots per circuit within the platform shot limit, we allocated a budget of $300{,}000$ shots per $\left(N, \alpha \right)$ pair per circuit over all randomized measurements.\footnote{We made an exception for the $N = 18$ non-stabilizerness measurements. The per-circuit shot count for $N = 18$ non-stabilizerness is $520{,}000$.\label{fn:N18}}

In~\cite{RM_probing_entanglement}, it was numerically demonstrated that, given a fixed shot budget, the optimal ratio of number of random measurements ($N_U$) to shots per random measurement ($N_M$), $\frac{N_U}{N_M}$, decreases exponentially in system size $N$ for estimating the purity of the reduced state on half of the qubits of a random pure state. Our final logical state is not Haar-random, but it is pure and random, highly dependent on the encoding circuit. We select a measurement scheme in which we decrease $N_U$ and increase $N_M$ as $N$ increases. Per-job platform limits prevent further optimization of this ratio.

It was also shown in~\cite{RM_probing_entanglement} that the total shot budget $N_U N_M$ required to achieve a fixed variance increases with system size $N$. $N_M$ is fixed for all $\alpha$ of a given $N$ and is applied before post-selection. The fraction of shots that are post-selected decreases with increasing $\alpha$ \cref{appendix_postselection}. These two facts mean that our standard deviations for our entanglement entropy estimators (and empirically, non-stabilizerness estimators) increase with increasing $\alpha$ and $N$.

\begin{table}[ht]\label{ee_meas_counts}
    \centering
    \caption{Entanglement entropy measurement settings for each system size.}
    \label{tab:rm-settings}
    \begin{tabular}{ccc}
        \toprule
        $N$ (system size) & \# RM & $N_M$ \\
        \midrule
        8  & 40 & 7{,}500  \\
        12 & 40  & 7{,}500  \\
        16 & 10  & 30{,}000 \\
        20 & 10   & 30{,}000 \\
        \bottomrule
    \end{tabular}
\end{table}

\begin{table}[ht]
    \centering
    \caption{Non-stabilizerness measurement settings for each system size.}
    \label{tab:rm-settings}
    \begin{tabular}{ccc}
        \toprule
        $N$ (system size) & \# RM & $N_M$ \\
        \midrule
        8  & 150 & 2{,}000  \\
        10 & 100 & 3{,}000  \\
        12 & 80  & 3{,}750  \\
        14 & 60  & 5{,}000  \\
        16 & 20  & 15{,}000 \\
        18 & 8   & 65{,}000 \\
        \bottomrule
    \end{tabular}
\end{table}

\section{Experimental Data}\label{experimental_data}

To restate, for each order parameter, we execute circuits with twelve $\alpha$ values: $\{0, 0.7, 0.8, 0.9, 0.933, 0.967, 1, 1.033, 1.067, 1.2, 1.3, \pi \}$. For each $(N, \alpha, i)$ triple, with $i$ indexing the order parameter, we sample four random encodings and execute each at three ZNE levels ($\lambda = 1,3,5$). Thus, for each $(N, i)$ pair, we execute $144$ distinct encoding-$\lambda$ circuits. For fidelity measurements, the shot count per each circuit is $30{,}000$. For entanglement entropy and non-stabilizerness, over the RM for each circuit, the shot count is $300{,}000$.\hyperref[fn:N18]{\footnotemark[\getrefnumber{fn:N18}]}
The total number of executed shots across all three order parameters, excluding M3 calibration jobs, is $5.0688 \times 10^8$.

\begin{figure*}[t]
    \centering
    \makebox[\textwidth][c]{%
    \includegraphics[width=0.32\textwidth]{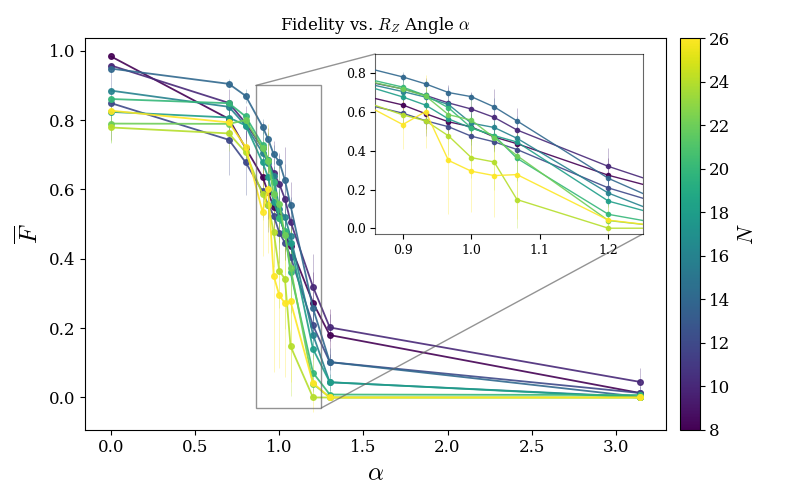}%
        \hspace{0.01\textwidth}%
    \includegraphics[width=0.32\textwidth]{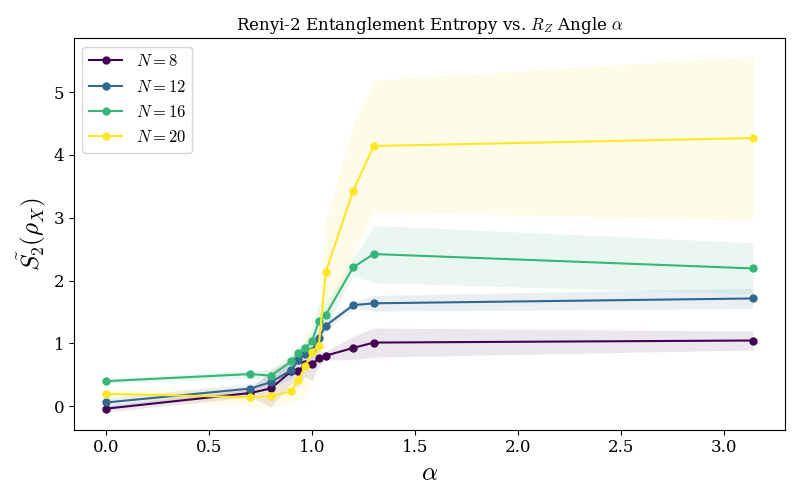}%
        \hspace{0.01\textwidth}%
    \includegraphics[width=0.32\textwidth]{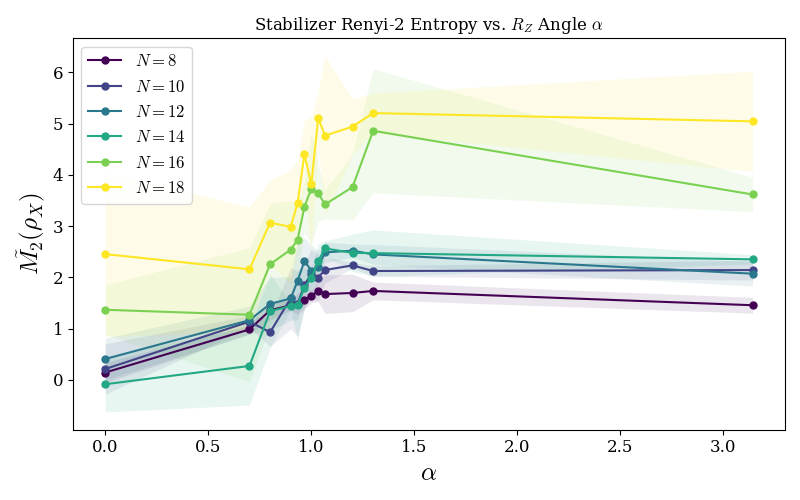}%
    }
    \caption{Error-mitigated experimental data without normalization by $\alpha=0$ values. \textbf{Left:} Fidelity vs.\ angle; inset zooms in on the critical region. \textbf{Center:} Rényi-2 entanglement entropy $\tilde{S}_2(\rho_X)$ vs.\ $\alpha$; shaded regions span one standard deviation. \textbf{Right:} Stabilizer Rényi-2 entropy $\tilde{M}_2(\rho_X)$ vs.\ $\alpha$.}
    \label{fig:raw_data_plots}
\end{figure*}

We plot the raw experimental data in \cref{fig:raw_data_plots}. For each order parameter, the error bars and shaded regions at each $(N, \alpha)$ display $\pm1$ standard error from the mean over the four encodings. Each encoding's estimator of each parameter is generated from finite-shot, error-mitigated data, meaning that the standard error reflects both inter-encoding variance and intra-encoding variance due to finite shot counts and error-mitigation noise. 

% \begin{figure*}[t]
%     \centering
%     \begin{subfigure}[b]{0.3\textwidth}
%         \centering
%         \includegraphics[width=\textwidth]{Figures/raw_fidelity_plot.png}
%         \caption{Plot of fidelity vs. angle raw data. Inset zooms in on critical region.}
%         \label{fig:raw_fid_plot}
%     \end{subfigure}
%     \hfill
%     \begin{subfigure}[b]{0.3\textwidth}
%         \centering
%         \includegraphics[width=\textwidth]{Figures/raw_ee_plot.png}
%         \caption{Joint posterior of $\alpha$ and $\nu^{(2)}$ for the $\tilde{S_2}(\rho_X)$ transition. Shaded regions span one standard deviation at each point.}
%         \label{fig:raw_ee_plot}
%     \end{subfigure}
%     \hfill
%     \begin{subfigure}[b]{0.3\textwidth}
%         \centering
%         \includegraphics[width=\textwidth]{Figures/raw_magic_plot.png}
%         \caption{Joint posterior of $\alpha$ and $\nu^{(3)}$ for the $\tilde{M_2}(\rho_X)$ transition.}
%         \label{fig:raw_magic_plot}
%     \end{subfigure}
%     \caption{Joint posteriors of $\alpha$ and critical exponent for each of the three order parameters.}
%     \label{fig:raw_data_plots}
% \end{figure*}

\subsection{Data ``Normalization"}\label{data_cleaning}

The finite size scaling functional forms in \cref{fidelity_scaling}, \cref{ee_scaling}, and \cref{magic_scaling} hold in the vicinity of $\alpha_c$. To avoid fitting our statistical model, described in \cref{GP_regression}, to regions in which such forms are not expected to hold, we omit $\alpha \in \{0, \pi\}$ data points from the model.

\begin{figure*}[t]
    \centering
    \makebox[\textwidth][c]{%
    \includegraphics[width=0.32\textwidth]{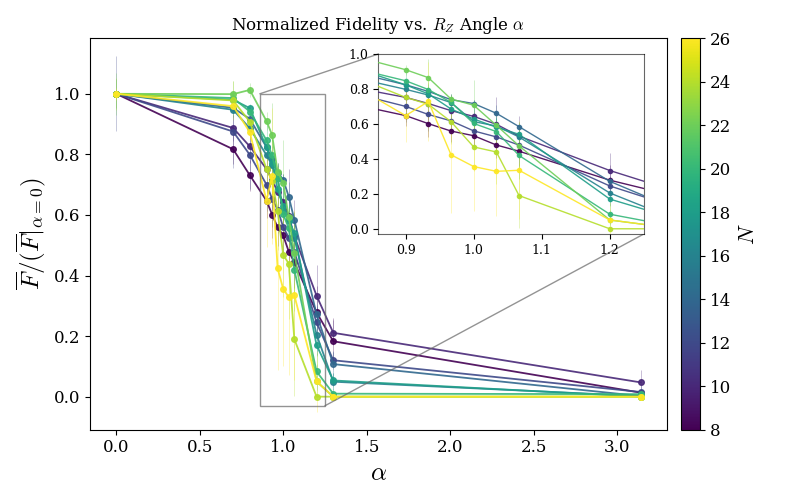}%
        \hspace{0.01\textwidth}%
    \includegraphics[width=0.32\textwidth]{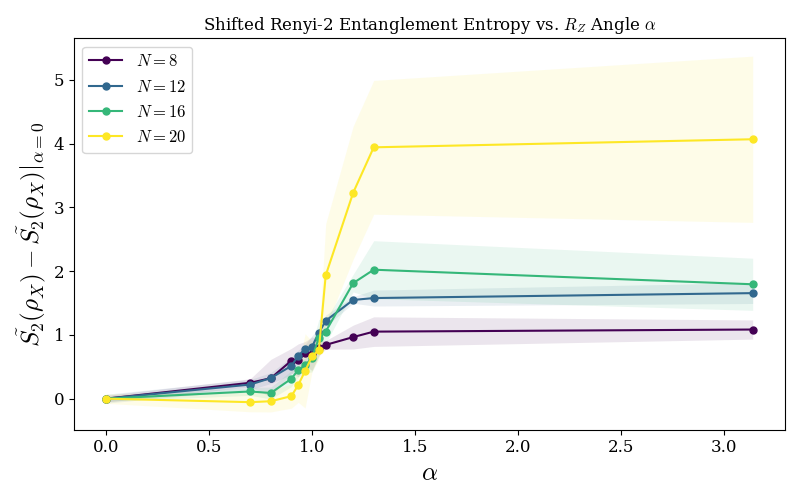}%
        \hspace{0.01\textwidth}%
    \includegraphics[width=0.32\textwidth]{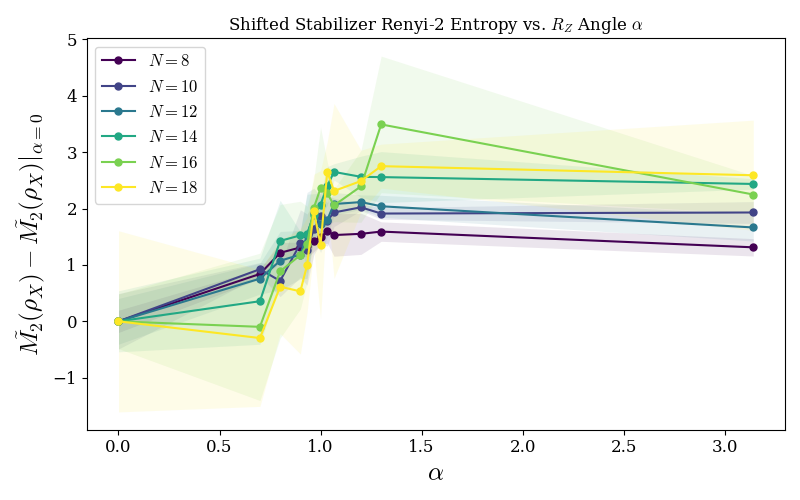}%
    }
    \caption{Normalized experimental data (by $\alpha = 0$ values). \textbf{Left:} Fidelity vs.\ angle; inset zooms in on the critical region. \textbf{Center:} Rényi-2 entanglement entropy $\tilde{S}_2(\rho_X)$ vs.\ $\alpha$; shaded regions span one standard deviation. \textbf{Right:} Stabilizer Rényi-2 entropy $\tilde{M}_2(\rho_X)$ vs.\ $\alpha$.}
    \label{fig:normalized_data_plots}
\end{figure*}

We do not want the decrease in fidelity as a function of $N$ (and corresponding increases in entanglement entropy and non-stabilizerness) to be spuriously incorporated into the critical parameter posteriors. We account for this by dividing each $\overline{F}$ data point by the $\alpha = 0 \ \overline{F}$ at the same $N$. Analogously, we subtract from each $\tilde{S_2}$ and $\tilde{M_2}$ point the values at $\alpha = 0$ for the same $N$. In other words, we normalize so that $\overline{F}$ at $\alpha = 0$ is $1$, and $\tilde{S_2}$ and $\tilde{M_2}$ at $\alpha = 0$ are $0$. This is effectively another layer of error mitigation. At a given $N$, each $\alpha$ just affects the virtual $R_Z$ layer. The attenuation of order parameters at $\alpha = 0$ is due to gate and readout noise that we expect to apply equally at each $\alpha$, and conveniently, at $\alpha = 0$ we know what the noiseless values of each order parameter would be. We plot the normalized order parameters in \cref{fig:normalized_data_plots}.

\section{Gaussian Process Regression}\label{GP_regression}

\subsection{Model Selection}

There are multitudinous methods to extract critical parameters via finite size scaling. We have three criteria we want our model of choice to satisfy. First, we want it to yield posterior distributions for the critical parameters. Given that there is no simple hypothesis test to detect the presence of a phase transition in this system, information about the distributions of the critical parameter estimates can indirectly convey confidence about the underlying supposition of a transition. Second, since each order parameter theoretically shares the same $\alpha_c$, we want a model that allows us to jointly use all three sets of experimental data to model the finite-size scaling. Third, we want a model expressive enough to capture the behavior of the functions in \cref{fidelity_scaling}, \cref{ee_scaling}, and \cref{magic_scaling}; for instance, a low-order polynomial expansion about $\alpha_c$, as done in~\cite{magic_phase_transition_experiment}, would not suffice.

Inspired by~\cite{GP_Harada}, we model the phase transition jointly over the three order parameters as three separate Gaussian processes (GPs), each sharing the same $\alpha_c$ and having its own $\nu^{(i)}$. Monte Carlo sampling from the posteriors of the critical parameters gives information about their distributions. Sharing $\alpha_c$ over the three GPs constrains our estimate of it using all of our experimental data. Finally, using a Gaussian process prior over each scaling collapse function allows us to avoid specifying a precise functional form of the scaling collapse function while constraining this function to be smooth and making inference tractable.

In the noiseless theory, each order parameter should share the same $\nu = 1$, as all three are functions of $\exp\left[\tan(\alpha_c/2)N(\alpha - \alpha_c) \right]$ near $\alpha_c$ \cref{theory}. We model each $\nu^{(i)}$ independently to account for finite-size effects. Since each order parameter is a different function of $\exp\left[\tan(\alpha_c/2)N(\alpha - \alpha_c) \right]$, they each have different sub-leading terms in their Taylor expansions. Our data span small values of $N$, which means their apparent $N$-dependence on this dataset will vary. Fixing a shared $\nu$ could lead to the model accounting for this by shifting $\alpha_c$. Additionally, allowing them to differ provides a test for the theoretical model of \cref{theory}: agreement between independent critical exponents is evidence for underlying universal behavior, rather than a forced assumption.

\subsection{Gaussian Process Regression Details}\label{GP_details}

Because our three measurements share the same critical angle $\alpha_c$, we use an explicit likelihood-based approach to finite size scaling estimation that estimates a single $\alpha_c$ by pooling information across order parameter measurement outcomes $\mathbf{Y}^{(1)}$, $\mathbf{Y}^{(2)}$, and $\mathbf{Y}^{(3)}$, corresponding to fidelity, R\'enyi-2 entanglement entropy, and stabilizer R\'enyi-2 entropy respectively. Here each $\mathbf{Y}^{(i)}$ is a $ n_i \times 10$ matrix, where for order parameter $i$, $\mathbf{Y}^{(i)}_{jk}$ is the measurement outcome for the $j^{\textrm{th}}$ system size and the $k^{\textrm{th}}$ $\alpha$ value $(\equiv \boldsymbol{\alpha}_k)$.  This requires jointly modeling the likelihood of $\mathbf{Y}^{(1)}$, $\mathbf{Y}^{(2)}$, and $\mathbf{Y}^{(3)}$ as a function of the critical angle $\alpha_c$ and critical exponents $\nu^{(1)}$, $\nu^{(2)}$, and $\nu^{(3)}$. Because our measurements of these phenomena are physically independent, we assume statistical independence of $\mathbf{Y}^{(1)}$, $\mathbf{Y}^{(2)}$, and $\mathbf{Y}^{(3)}$ \textit{conditional} on the critical parameters and scaling collapse functions. That is, we assume
\begin{equation*}
    p(\mathbf{Y}^{(1)}, \mathbf{Y}^{(2)}, \mathbf{Y}^{(3)} \mid  \alpha_c, \boldsymbol{\Theta}) = \prod_{i = 1}^3 p(\mathbf{Y}^{(i)} \mid  \alpha_c, \boldsymbol{\theta}^{(i)})
\end{equation*}
where $\boldsymbol{\Theta} = (\boldsymbol{\theta}^{(1)}, \boldsymbol{\theta}^{(2)}, \boldsymbol{\theta}^{(3)})$, and each  $\boldsymbol{\theta}^{(i)} = (\nu^{(i)}, f^{(i)},\beta^{(i)}, \eta^{(i)})$, where $f^{(i)}$ is the scaling collapse function for observable $i$, and $\beta^{(i)}, \eta^{(i)}$ are additional parameters necessary for inference but without a physical interpretation, as explained below.

To arrive at a likelihood for the measurements of the individual observables $ p(\mathbf{Y}^{(i)} \mid \alpha_c, \boldsymbol{\theta}^{(i)})$, we note that our observations over system sizes $N_j$ are physically independent. Letting $\phi(x; \mu, \sigma^2)$ denote the density of the normal distribution with mean $\mu$ and variance $\sigma^2$ evaluated at $x$, we assume
\begin{equation*}
    p(\mathbf{Y}^{(i)} \mid \alpha_c, \boldsymbol{\theta}^{(i)}) = \prod_{j = 1}^{n_i}\prod_{k = 1}^{10} \phi\left(\mathbf{Y}^{(i)}_{jk} ; \mathbf{M}_{jk}^{(i)}, \boldsymbol{\Sigma}^{(i)}_{jk}\right),
\end{equation*}
with 
\begin{align*}
    \mathbf{M}_{jk}^{(i)} &= \beta^{(i)} +  f^{(i)}\left((\boldsymbol{\alpha}_k - \alpha_c)N_j^{1/\nu^{(i)}}\right) \\
    \boldsymbol{\Sigma}^{(i)}_{jk} &= (\mathbf{S}^{(i)}_{jk})^2 + (\eta^{(i)})^2,
\end{align*}
where $\beta^{(i)}$ is an intercept term that shifts the location of the curve, $\mathbf{S}^{(i)}_{jk}$ is our estimate of the empirical standard error of $\mathbf{Y}^{(i)}_{jk}$, and $(\eta^{(i)})^2$ is an additional variance parameter that accounts for the fact that the scaling collapse ansatz holds only asymptotically, so even noiseless data would not fall exactly on the scaling collapse curve. Here, we have assumed a normal distribution for modeling simplicity; we provide numerical justification for this in \cref{normality_numerics}.

While the scaling collapse functions $f^{(i)}$ have been derived theoretically in~\cite{enc_dec_original, enc_dec_magic}, our present goal is to infer \textit{if} the empirical data conform to these theoretical results. Thus, instead of assuming a particular functional form of the scaling collapse functions, we instead model each $f^{(i)}$ as an unknown random variable, which allows us to reflect our \textit{a posteriori} uncertainty about the form of the scaling collapse function given the observable data. As argued in~\cite{GP_Harada}, we expect each scaling collapse function to be smooth, and thus we assume as a prior distribution that each $f^{(i)}$ is drawn from a mean $0$ Gaussian process with a squared-exponential covariance that decays with the squared distance between scaled data points. That is, we assume
\begin{equation*}
    f^{(i)} \sim \mathcal{GP}(\mathbf{0}, \mathbf{C}^{(i)}(\cdot,\cdot)),
\end{equation*}
where 
\begin{equation*}
    \mathbf{C}^{(i)}(\mathbf{X}^{(i)}_{jk},\mathbf{X}^{(i)}_{mp}) = (\sigma^{(i)})^2 \exp\left(-\frac{(\mathbf{X}^{(i)}_{jk} - \mathbf{X}^{(i)}_{mp})^2}{2(\ell^{(i)})^2}\right)
\end{equation*}
expresses the covariance between points $\mathbf{X}^{(i)}_{jk} = (\boldsymbol{\alpha}_k - \alpha_c)N_j^{1/\nu^{(i)}}$ and $\mathbf{X}^{(i)}_{mp} = (\boldsymbol{\alpha}_p - \alpha_c)N_m^{1/\nu^{(i)}}$. We further assume priors on the marginal standard deviation $\sigma^{(i)}$ and lengthscale $\ell^{(i)}$ for each scaling collapse function $i$ as well as on the critical parameters themselves as described in the appendix. 

Given our measurements $\mathbf{Y}^{(1)}$, $\mathbf{Y}^{(2)}$, and $\mathbf{Y}^{(3)}$, we
wish to infer a joint posterior distribution over the scaling collapse functions
$f^{(1)}$, $f^{(2)}$, and $f^{(3)}$ together with the critical parameters $\alpha_c$,
$\nu^{(1)}$, $\nu^{(2)}$, and $\nu^{(3)}$. While the scaling collapse functions themselves are infinite-dimensional objects, practically we must restrict our sampling over the posterior to a finite set of points. Since each Gaussian process enters the likelihood only
through its values at the observed scaled locations
$\mathbf{X}^{(i)}$, we choose to evaluate the posteriors over the scaling collapse functions only at the $n_i \times 10$ matrices $\mathbf{F}^{(i)}$, where 
\begin{equation*}
    \mathbf{F}^{(i)}_{jk} = f^{(i)}(\mathbf{X}^{(i)}_{jk}).
\end{equation*}
While the resulting posterior is analytically intractable, we draw
approximate samples from it using the No-U-Turn Sampler~\cite{NUTS}. 

Additional details on the Gaussian process are given in the appendix, along with Table \ref{tab:symbols} that describes the notation of this section.

\section{GP Regression Results and Evidence for Phase Transition}\label{GP_regression_results}

\begin{figure*}[t]
    \centering
    \includegraphics[width=0.9\textwidth]{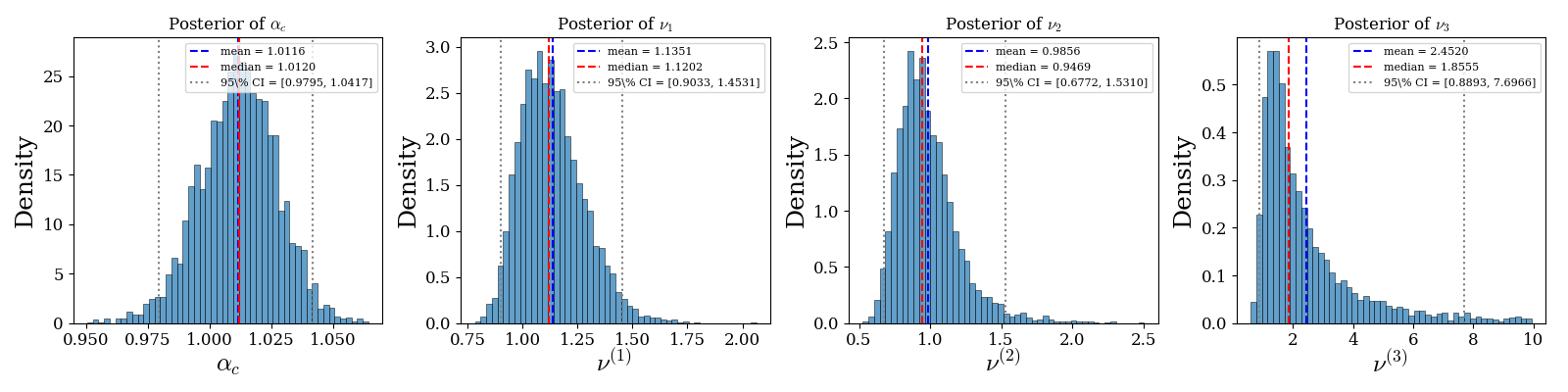}
    \caption{Histograms reflecting posteriors of the critical parameters $\alpha_c$, $\nu^{(1)}$ (fidelity), $\nu^{(2)}$ (entanglement entropy), $\nu^{(3)}$ (non-stabilizerness). Each is generated from 1,000 samples.}
    \label{fig:crit_param_posteriors}
\end{figure*}

We display histograms of the critical parameter posteriors in \cref{fig:crit_param_posteriors}. The critical angle distribution $\alpha_c$ is concentrated tightly around $1.011$, indicating the model is ``confident" about that value. The relative variance of the posteriors increases from $\alpha_c$ to $\nu^{(3)}$. This behavior is expected. Each order parameter tightens the estimate of the shared parameter $\alpha_c$. Furthermore, fidelity, entanglement entropy, and non-stabilizerness depend on ascending moments of the underlying state $\rho_X$, with each successive order parameter thereby requiring more measurement shots to achieve the same relative variance. Our experimental data reflect our constrained shot budget, providing greater confidence for lower-moment order-parameter critical exponents.

\begin{figure*}[t]
    \centering
    \begin{subfigure}[b]{0.3\textwidth}
        \centering
        \includegraphics[width=\textwidth]{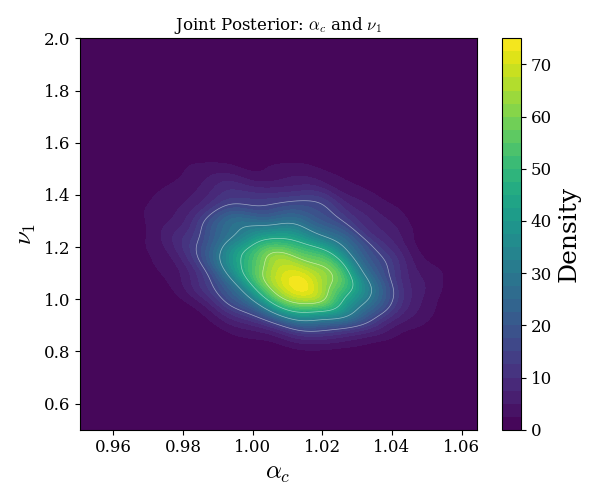}
        \caption{Joint posterior of $\alpha_c$ and $\nu^{(1)}$ for the $\overline{F}$ transition.}
        \label{fig:joint_posterior_nu1}
    \end{subfigure}
    \hfill
    \begin{subfigure}[b]{0.3\textwidth}
        \centering
        \includegraphics[width=\textwidth]{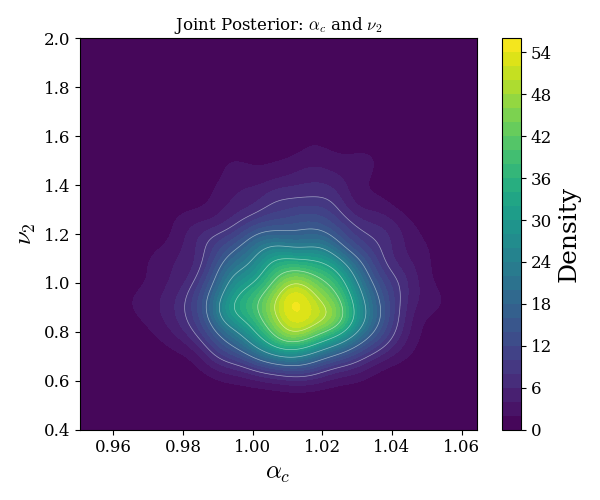}
        \caption{Joint posterior of $\alpha_c$ and $\nu^{(2)}$ for the $\tilde{S_2}(\rho_X)$ transition.}
        \label{fig:joint_posterior_nu2}
    \end{subfigure}
    \hfill
    \begin{subfigure}[b]{0.3\textwidth}
        \centering
        \includegraphics[width=\textwidth]{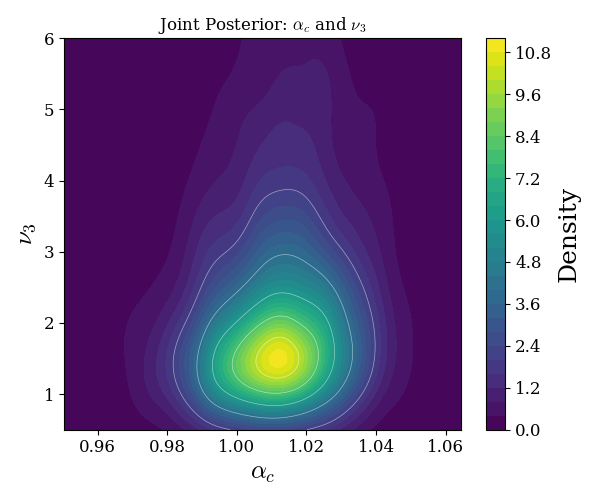}
        \caption{Joint posterior of $\alpha_c$ and $\nu^{(3)}$ for the $\tilde{M_2}(\rho_X)$ transition.}
        \label{fig:joint_posterior_nu3}
    \end{subfigure}
    \caption{Joint posteriors of $\alpha_c$ and critical exponent for each of the three order parameters.}
    \label{fig:joint_posteriors}
\end{figure*}

In \cref{fig:joint_posteriors}, we display the joint posteriors for each of the three order parameters. For each, the semi-major and semi-minor axes of the high-density ellipses are reasonably well-aligned with the critical parameter axes, indicating that the values of $\alpha_c$ and $\nu^{(i)}$ are not strongly correlated for any $i$. This implies that our model is not spuriously constraining $\alpha_c$ to account for underdetermined critical exponents.

\begin{figure*}[t]
    \centering
    \begin{subfigure}[b]{0.3\textwidth}
        \centering
        \includegraphics[width=\textwidth]{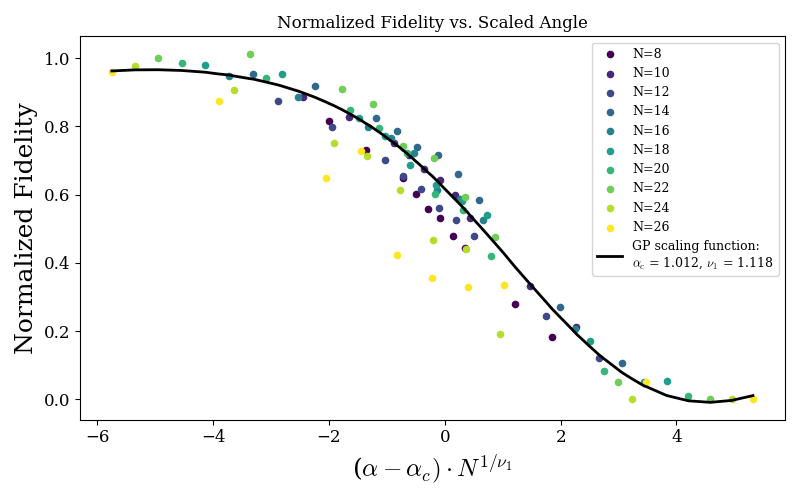}
        \caption{Scaling collapse for the normalized fidelity.}
        \label{fig:scaled_collapse_fid}
    \end{subfigure}
    \hfill
    \begin{subfigure}[b]{0.3\textwidth}
        \centering
        \includegraphics[width=\textwidth]{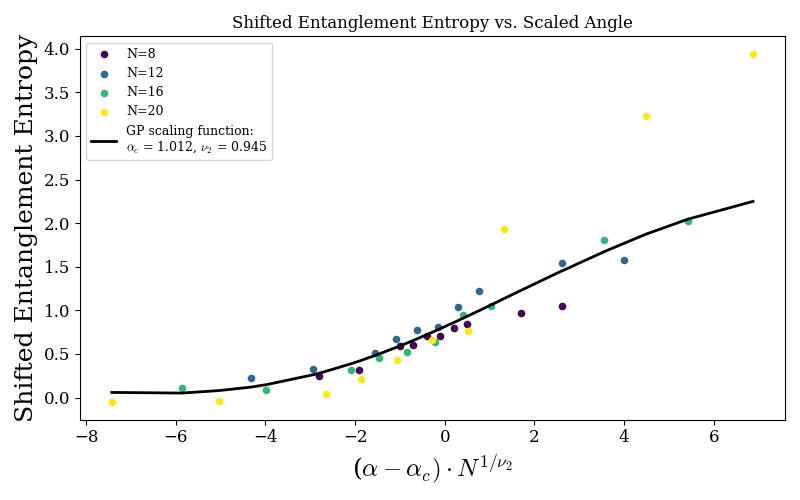}
        \caption{Scaling collapse for shifted R\'enyi-2 entanglement entropy.}
        \label{fig:scaled_collapse_ee}
    \end{subfigure}
    \hfill
    \begin{subfigure}[b]{0.3\textwidth}
        \centering
        \includegraphics[width=\textwidth]{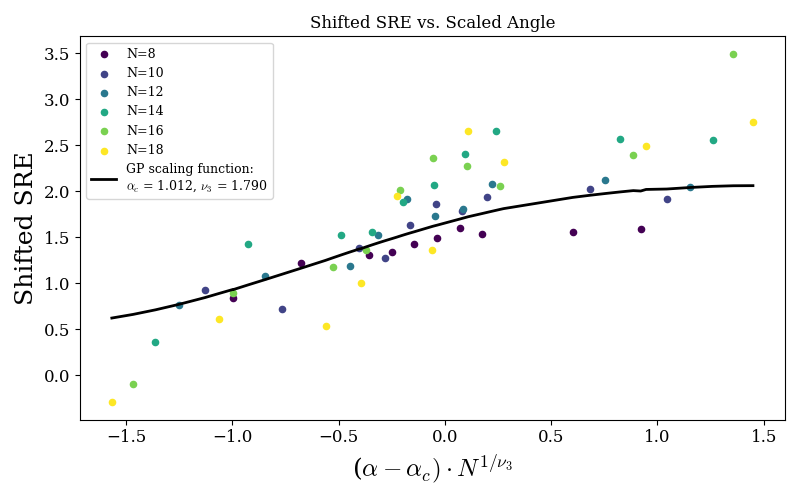}
        \caption{Scaling collapse for shifted stabilizer R\'enyi-2 entropy.}
        \label{fig:scaled_collapse_mag}
    \end{subfigure}
    \caption{Scaling collapses for experimental data of all three order parameters.}
    \label{fig:scaled_collapses}
\end{figure*}

We plot the scaled data collapses in \cref{fig:scaled_collapses}. The black curves are the GP means at each scaled coordinate. In each plot, the largest-$N$ points exhibit the biggest deviations from the GP curves. This is less a reflection of parameter drift in $N$ than of the fact that raw data variances grow strongly in $N$. Since the total shot budget per $(N, \alpha)$ pair is fixed and the post-selection probability decays strongly in $N$ \cref{appendix_postselection}, the total number of post-selected shots also decays strongly in $N$, and the variance thereby increases. The GP ``penalizes" deviations at low-variance points more, so it is expected that the largest deviations in the scaled collapse curves correspond to the highest-variance data.

\subsection{Analysis of Critical Parameters}

\begin{figure*}[t]
    \centering
    \includegraphics[width=0.9\textwidth]{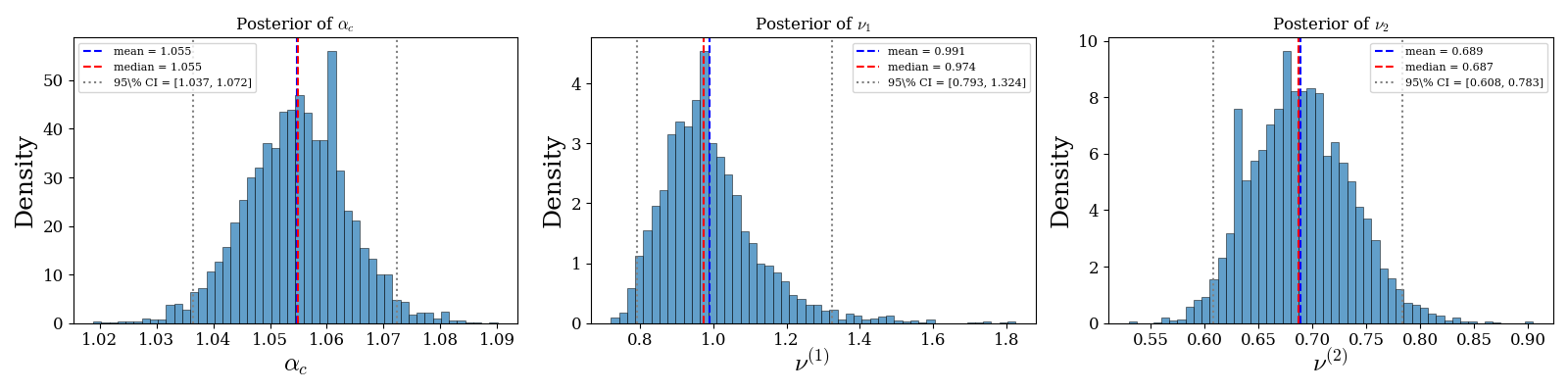}
    \caption{Histograms reflecting posteriors of the critical parameters $\alpha_c$, $\nu^{(1)}$ (fidelity), $\nu^{(2)}$ (entanglement entropy) from simulated data. Each is generated from 1,000 samples.}
    \label{fig:sim_crit_param_posteriors}
\end{figure*}

We observe a deviation in $\alpha_c$ between the posterior in \cref{fig:crit_param_posteriors} and the theoretical result in \cref{theory}. It has two main potential sources: first, differences between our random circuit ensemble and the Haar measure, and second, real device noise. To accurately determine the extent to which each source explains the discrepancy, we look to the simulated data in \cref{fig:sim_fid_and_ee}. Plugging the simulated data into a joint Gaussian process regression for fidelity and entanglement entropy, using the same priors as the experimental data model, yields the posteriors in \cref{fig:sim_crit_param_posteriors}. The simulated $\alpha_c$ falls very close to the experimental $\alpha_c$, with overlapping $95\%$ credible intervals. Both $\nu^{(1)}$ estimates fall near $1$, and the $\nu^{(2)}$ $95\%$ credible intervals overlap. These results indicate that after error mitigation, real device noise does not substantially shift the critical parameters.

The circuit ensemble therefore seems to account for the lower $\alpha_c$. This has a clear intuitive explanation: at low $N$ values, our circuits use a limited single-qubit gate family and are not deep enough to be truly Haar-random, thereby generally scrambling information less strongly. Information scrambling counteracts the effect of local noise in this system, so less scrambling leads to a lower threshold at which noise destroys information recoverability, corresponding to a lower critical angle.

In the absence of an obvious hypothesis test for the existence of a phase transition in our experiment, the above results provide evidence for one. The Gaussian process does indeed assume the form of the scaling ansatz $O^{(i)} = f(N^{1/\nu^{(i)}}(\alpha-\alpha_c))$. However, the visual collapse of the experimental data in \cref{fig:scaled_collapses} onto roughly single curves, the agreement between the critical-parameter posteriors of simulated and experimental data, and the tightness of the posterior distributions are all phenomena that would not be expected to manifest in the absence of transition-like behavior. These jointly lend credence to the ansatz itself and thereby to the existence of a transition in our circuit ensemble.

\subsection{Effect of Error Mitigation on Phase Transition Observation}

We have seen in \cref{error_mitigation} that error mitigation substantially increases the logical state fidelity at $\alpha = 0$. We now see that error mitigation is crucial for observing the error-resilience transition at all.

\begin{figure*}[t]
    \centering
    \includegraphics[width=0.9\textwidth]{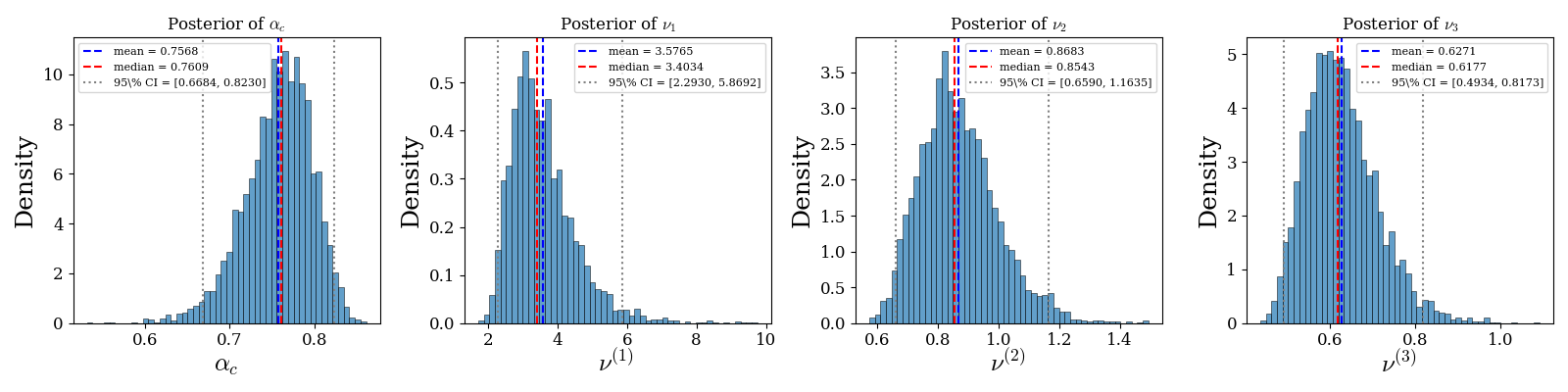}
    \caption{Histograms reflecting posteriors of the critical parameters $\alpha_c$, $\nu^{(1)}$ (fidelity), $\nu^{(2)}$ (entanglement entropy), $\nu^{(3)}$ (non-stabilizerness) from unmitigated experimental data. Each is generated from 1,000 samples.}
    \label{fig:unmitigated_crit_param_posteriors}
\end{figure*}

\begin{figure*}[t]
    \centering
    \begin{subfigure}[b]{0.3\textwidth}
        \centering
        \includegraphics[width=\textwidth]{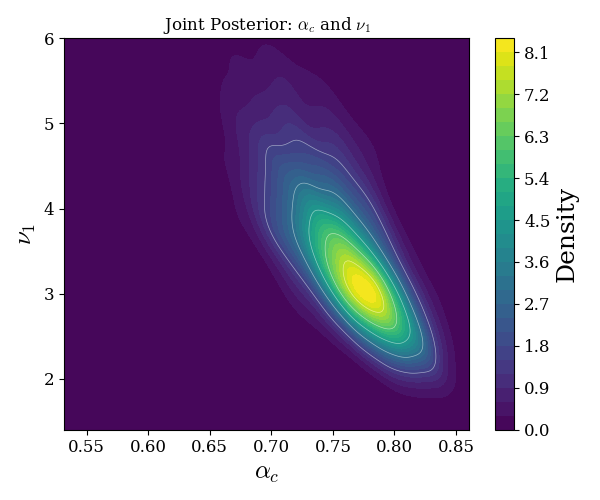}
        \caption{Joint posterior of $\alpha_c$ and $\nu^{(1)}$ for the $\overline{F}$ transition.}
        \label{fig:joint_posterior_nu1_unmitigated}
    \end{subfigure}
    \hfill
    \begin{subfigure}[b]{0.3\textwidth}
        \centering
        \includegraphics[width=\textwidth]{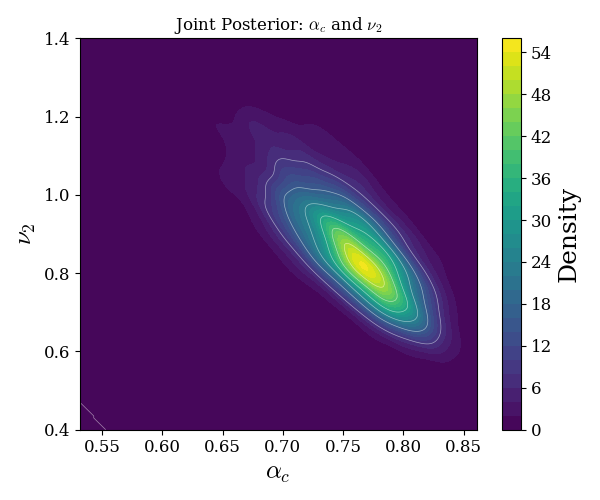}
        \caption{Joint posterior of $\alpha_c$ and $\nu^{(2)}$ for the $\tilde{S_2}(\rho_X)$ transition.}
        \label{fig:joint_posterior_nu2_unmitigated}
    \end{subfigure}
    \hfill
    \begin{subfigure}[b]{0.3\textwidth}
        \centering
        \includegraphics[width=\textwidth]{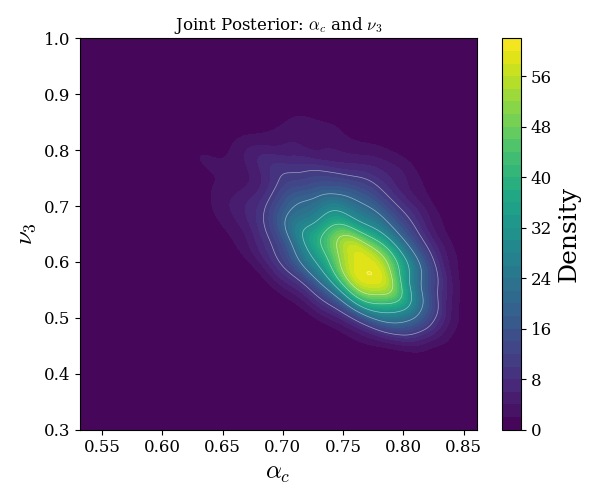}
        \caption{Joint posterior of $\alpha_c$ and $\nu^{(3)}$ for the $\tilde{M_2}(\rho_X)$ transition.}
        \label{fig:joint_posterior_nu3_unmitigated}
    \end{subfigure}
    \caption{Joint posteriors of $\alpha_c$ and critical exponent for each of the three order parameters using unmitigated experimental data.}
    \label{fig:joint_posteriors_unmitigated}
\end{figure*}

To demonstrate this, we take unmitigated experimental data, apply to them the same ``normalization" as described in \cref{data_cleaning}, and input them into the same Gaussian process model as the mitigated data. The individual posteriors are plotted in \cref{fig:unmitigated_crit_param_posteriors}, and the joint posteriors for each order parameter are plotted in \cref{fig:joint_posteriors_unmitigated}.

The model is unable to accurately identify $\alpha_c$, and its mean deviates strongly from both simulated and error-mitigated results. The $\alpha_c$ credible interval is far wider as well, indicating the model's inability to constrain the parameter. The relative lack of confidence in $\nu^{(1)}$ compared to $\nu^{(2)}$ and $\nu^{(3)}$, even though there are far more fidelity data than there are for other order parameters, indicates the model is struggling to jointly fit the critical parameters. This is further evident in the joint posterior plots of \cref{fig:joint_posteriors_unmitigated}, whose noticeably diagonal high-density regions for each order parameter indicate the model is determining $\alpha_c$ to constrain the underdetermined critical exponents. Overall, these observations point to a poor fit and do not give credence to genuine critical behavior.

One possible objection is that the model priors are to blame for these posteriors. However, the strong consistency in $\alpha_c$ between the simulated and error-mitigated data, and visual consistency of the order parameter plots with $\alpha_c$, indicates that the priors are set reasonably. It appears far more likely that error mitigation allows the statistical model to accurately identify the critical behavior. We have accounted for $N$-dependent noise in offsetting by the $\alpha = 0$ order parameter values, but the effects of device noise display an $\alpha$-dependence that grows more pernicious as $N$ increases, leading larger-$N$ curves to steepen in the critical region more quickly than smaller ones. Intuitively device noise both affects baseline fidelity, entropy, and non-stabilizerness, while also reducing the error-resilience of the circuit ensembles, phenomena that intensify as $N$ increases. This prevents the order-parameter curves from crossing at the critical point, as described in \cref{theory} ($\overline{F} = 1/2, \overline{S_2}(\rho_{|X_1|}) = 2, \overline{M_2}(\rho) = 4$). Error mitigation ameliorates some of this, bringing the observed curves closer together in the critical region.

\subsection{Evidence for Observation of Phase Transition}

We now summarize the evidence that our experiment displays signatures of the error-resilience phase transition. First, we have a strong prior expectation that our hardware circuits would display such critical behavior. The theoretical model in \cref{theory} has an error-resilience transition, and our experimental, ``approximate" circuit model \cref{experiment_implement} displays similar behavior, appearing to converge to the same $\alpha_c$ as $N \rightarrow \infty$. The experimental question is whether or not device noise undermines this phenomenon.

Our foremost experimental evidence is the set of plots in \cref{fig:normalized_data_plots}. The fidelity and entanglement entropy plots both display behavior expected for finite-sized circuit ensembles whose large-$N$ behavior displays the transition in \cref{theory}, while the statistically much noisier non-stabilizerness plot is consistent with it. Namely, we see plots of ascending $N$ values roughly crossing each other at similar $\alpha$ values, as expected via \cref{theory} and \cref{experiment_implement}. 

We classically simulate the noiseless transition using the same circuit ensemble as our device experiments. We then pass our classical simulation data, error-mitigated experimental data, and unmitigated experimental data through the same statistical model. We observe that the simulated and error-mitigated experimental data yield similar effective $\alpha_c$ and $\nu^{(1)}$ values. This provides numerical evidence that the error-mitigated experimental data captures the predicted transition. The shape of the marginal $(\alpha_c, \nu)$ posteriors further bolsters our confidence in the statistical fit, as it indicates that the model is able to fairly independently constrain the parameter without them compensating against each other. Independence reflects the model's capacity to precisely pinpoint the critical behavior rather than forcing the ansatz to hold by trading off values in a degenerate parameter space. 

One may object that the above evidence presupposes the validity of the finite-size scaling ansatz and priors of the GP model. The unmitigated experimental data addresses this concern by acting as a sort of negative control. By a wide margin, it fails to converge to posteriors matching the simulated data, indicating that our model does not ``force" the effective critical parameters to the simulated values.

We reiterate that the effective $\alpha_c$ we observe experimentally differs from the $\alpha_c$ in the Haar-random model of \cref{theory} and the likely large-$N$ $\alpha_c$ of our experimentally implemented circuit ensemble. We believe that this ensemble approaches the behavior of the Haar-random model as $N$ increases, which the simulations seem to indicate. However, our observed $\alpha_c$ reflects the behavior of this ensemble at small $N$ values, consistent with simulation. Thus, while not capturing the exact $\alpha_c$ of the thermodynamic limit, our experiments display empirical signatures of the transition.

\section{Conclusion}

In this work, we have demonstrated the first experimental observation of the error-resilience phase transition in noisy encoding-decoding circuits described in~\cite{enc_dec_original, enc_dec_magic}. Using IBM's \texttt{ibm\_pittsburgh} Heron r3 device, we have measured fidelity, entanglement entropy, and non-stabilizerness in systems of up to $N=26$ qubits. Detecting meaningful signal in the latter two quantities has required the use of single-copy randomized measurement protocols, variance reduction methods, and large shot budgets. This experimental overhead allows our Gaussian process model to identify critical parameters for finite-size scaling data collapse.

We observe a critical angle $\alpha_c$ of $\sim 1.01$ as opposed to the Haar-random $\alpha_c \approx 1.14$. To understand this deviation, we simulate noiseless encoding-decoding circuits from the same random circuit ensemble as in our experiments and input the data into the same Gaussian process regression. The simulated data yields $\alpha_c \approx 1.05$, indicating close agreement between simulation and experiment. This locates the source of the deviation in the encoding circuit ensemble, which generates circuits not deep enough to truly be Haar-random, which thereby scramble less and are less noise-resilient. With error mitigation, transition is therefore seen to be robust enough to persist in spite of real device noise.

We also see that our use of the error mitigation methods ZNE and M3 is essential for detecting critical behavior. The Gaussian process regression applied to unmitigated experimental data gives critical parameter posteriors that are highly correlated and inconsistent with the observed order parameter curves, indicating a failure to identify criticality.

Our model most tightly constrains the fidelity critical parameters, followed by entanglement entropy, then non-stabilizerness. As order parameters depend on increasing moments $n$ of the state ($\rho_X^{\otimes n})$, a greater experimental budget is required to constrain their critical parameters. Across all three, our data are consistent with and informative about the same error-resilience transition. However, to further constrain the critical parameters given a finite resource budget, further refinement of the techniques used in this work is required. For instance, strategically allocating our total shot budget \textit{across} all three order parameters could increase our credence in the overall error-resilience transition. Furthermore, more precise optimization over the tradeoff between number of random measurements and shots per measurement, as well as better approximate state ans\"atze, could significantly reduce the variance of our estimators.

Nonetheless, the suite of techniques deployed in conjunction can inform experimental design and data processing for the observation of critical phenomena on NISQ-era devices. We incorporate experimental data into our variance reduction method scalably, apply error mitigation, employ a statistical model that allows joint finite-size scaling collapse across order parameters, and benchmark our results against simulation and raw data. These methods extend easily to the investigation of other information-theoretic phase transitions, enabling more extensive experiments on NISQ-era devices.

\section{Appendix - GP}\label{Appendix_GP}

\subsection{Model Specification}

While the scaling collapse functions $f^{(1)}, f^{(2)},$ and $f^{(3)}$ are infinite-dimensional objects, the likelihood of our observables only depends upon the values of these functions at the finite set of points for which we observe the (scaled) data. Collecting these values into the $n_i \times 10$ matrices $\mathbf{F}^{(i)}$ with $\mathbf{F}^{(i)}_{jk} = f^{(i)}(\mathbf{X}^{(i)}_{jk})$, and writing $\vec{\mathbf{F}}^{(i)} = \operatorname{vec}(\mathbf{F}^{(i)})$ for the vector obtained by stacking the entries of $\mathbf{F}^{(i)}$ by column, we can equivalently write our model as
\begin{align*}
    \mathbf{Y}^{(i)}_{jk} \mid 
    \mathbf{M}^{(i)}_{jk}, \boldsymbol{\Sigma}^{(i)}_{jk}
        &\sim \mathcal{N}\!\left(\mathbf{M}^{(i)}_{jk},\; \boldsymbol{\Sigma}^{(i)}_{jk}\right) \\
        \mathbf{M}^{(i)}_{jk} &= \beta^{(i)} + \mathbf{F}^{(i)}_{jk} \\
        \boldsymbol{\Sigma}^{(i)}_{jk} &= (\mathbf{S}^{(i)}_{jk})^2 + {\left(\eta^{(i)}\right)}^2\\
    \vec{\mathbf{F}}^{(i)} \mid \widetilde{\mathbf{C}}^{(i)}
        &\sim \mathcal{N}\!\left(\mathbf{0},\; \widetilde{\mathbf{C}}^{(i)}\right), \\
    \widetilde{\mathbf{C}}^{(i)}_{(jk),(mp)}
        &= (\sigma^{(i)})^2 \exp\!\left(-\frac{(\mathbf{X}^{(i)}_{jk} - \mathbf{X}^{(i)}_{mp})^2}{2(\ell^{(i)})^2}\right), \\
    \mathbf{X}^{(i)}_{jk} &= (\boldsymbol{\alpha}_k - \alpha_c)\,N_j^{1/\nu^{(i)}}, \\
    \beta^{(1)}, \beta^{(2)}, \beta^{(3)} &\sim p(\beta^{(i)}), \\
    \sigma^{(1)}, \sigma^{(2)}, \sigma^{(3)} &\sim p(\sigma^{(i)}), \\
    \ell^{(1)}, \ell^{(2)}, \ell^{(3)} &\sim p(\ell^{(i)}), \\
    \eta^{(1)}, \eta^{(2)}, \eta^{(3)} &\sim p(\eta^{(i)}), \\
    c^{(i)} &= 1 / \nu^{(i)} \\
    c^{(1)}, c^{(2)}, c^{(3)} &\sim p(c^{(i)}), \\
    \alpha_c &\sim p(\alpha_c),
\end{align*}
where the priors for each of these terms are specified in Table \ref{tab:priors}.
\begin{table*}[t]
\centering
\caption{Summary of symbols used in the finite-size-scaling model. }
\label{tab:symbols}
\begin{tabular}{lll}
\hline
Symbol & Type & Description \\
\hline
\multicolumn{3}{l}{\textit{Indices and data}} \\
$i$ & index & Observable label, $i \in \{1,2,3\}$ \\
$j$ & index & System-size label, $j \in \{1,\dots,n_i\}$ \\
$k$ & index & Angle-value label, $k \in \{1,\dots,10\}$ \\
$n_i$ & integer & Number of system sizes available for observable $i$ \\
$N_j$ & scalar & Linear system size for row $j$ \\
$\mathbf{Y}^{(i)}$ & $n_i \times 10$ matrix & Measured outcomes for observable $i$ \\
$\mathbf{Y}^{(i)}_{jk}$ & scalar & Measurement at system size $j$, angle $k$ \\
$\boldsymbol{\alpha}_k$ & scalar & $k$th angle value on the (shared) grid, excluding $0$ and $\pi$ \\
$\mathbf{S}^{(i)}_{jk}$ & scalar & Empirical standard error of $\mathbf{Y}^{(i)}_{jk}$ \\
\hline
\multicolumn{3}{l}{\textit{Critical parameters (of physical interest)}} \\
$\alpha_c$ & scalar & Critical angle, shared across all observables \\
$\nu^{(i)}$ & scalar & Critical exponent for observable $i$ \\
$c^{(i)}$ & scalar & Sampled reparametrization of the exponent, $c^{(i)} = 1/\nu^{(i)}$ \\
\hline
\multicolumn{3}{l}{\textit{Scaling-collapse functions}} \\
$f^{(i)}$ & function & Scaling-collapse function for observable $i$ \\
$\mathbf{X}^{(i)}_{jk}$ & scalar & Scaled coordinate $(\boldsymbol{\alpha_k} - \alpha_c)N_j^{1/\nu^{(i)}}$ \\
$\mathbf{F}^{(i)}$ & $n_i \times 10$ matrix & Values of $f^{(i)}(\mathbf{X}^{(i)}_{jk})$ \\
$\Vec{\mathbf{F}}^{(i)}$ & length $10n_i$ vector & Values of $\mathbf{F}^{(i)}$ stacked by column \\
\hline
\multicolumn{3}{l}{\textit{Likelihood}} \\
$\mathbf{M}^{(i)}_{jk}$ & scalar & Likelihood mean, $\beta^{(i)} + \mathbf{F}^{(i)}_{jk}$ \\
$\boldsymbol{\Sigma}^{(i)}_{jk}$ & scalar & Likelihood variance, $(\mathbf{S}^{(i)}_{jk})^2 + (\eta^{(i)})^2$ \\
$\boldsymbol{\theta}^{(i)}$ & collection & Parameters for observable $i$, $(\nu^{(i)}, f^{(i)}, \beta^{(i)}, \eta^{(i)})$ \\
$\boldsymbol{\Theta}$ & collection & All observables' parameters, $(\boldsymbol{\theta}^{(1)}, \boldsymbol{\theta}^{(2)}, \boldsymbol{\theta}^{(3)})$ \\
\hline
\multicolumn{3}{l}{\textit{Nuisance hyper-parameters}} \\
$\beta^{(i)}$ & scalar & Intercept shifting the collapse curve vertically \\
$\eta^{(i)}$ & scalar & Excess-noise std.\ dev.\  \\
$\sigma^{(i)}$ & scalar & GP marginal standard deviation  \\
$\ell^{(i)}$ & scalar & GP lengthscale \\
$\mathbf{C}^{(i)}$ & function & GP kernel for observable $i$ \\
$\widetilde{\mathbf{C}}^{(i)}$ & $10n_i \times 10n_i$ matrix & Covariance of $\Vec{\mathbf{F}}^{(i)}$ induced by $\mathbf{C}^{(i)}$ \\
\hline
\end{tabular}
\end{table*}
\begin{table*}[t]
\centering
\caption{Prior distributions for all model parameters. The index $i \in \{1,2,3\}$
runs over the three observables (fidelity, R\'enyi-2 entanglement entropy, stabilizer R\'enyi-2 entropy).
$\mathcal{N}(\mu,\sigma^2)$ denotes a normal distribution;
$\mathcal{N}^+$ a half-normal (a normal truncated to the positive half-line, further restricted to the support listed);
$\mathcal{N}_{[a,b]}$ a normal truncated to $[a,b]$;
$\mathrm{LogNormal}(\mu,\sigma^2)$ a log-normal whose logarithm has mean $\mu$ and variance $\sigma^2$.}
\label{tab:priors}
\begin{tabular}{llll}
\hline
Parameter & Description & Support & Prior \\
\hline
\multicolumn{4}{l}{\textit{Shared critical parameter}} \\
$\alpha_c$        & Critical angle (shared)            & $\mathbb{R}$        & $\mathcal{N}(1,\,0.5^2)$ \\
\hline
\multicolumn{4}{l}{\textit{Critical exponents} ($c^{(i)} = 1/\nu^{(i)}$)} \\
$c^{(1)}$ & Fidelity exponent           & $[0.1,\,3]$ & $\mathcal{N}_{[0.1,3]}(0.75,\,0.5^2)$ \\
$c^{(2)}$ & Entropy exponent            & $[0.1,\,3]$ & $\mathcal{N}_{[0.1,3]}(0.75,\,0.5^2)$ \\
$c^{(3)}$ & Magic exponent              & $[0.1,\,3]$ & $\mathcal{N}_{[0.1,3]}(0.75,\,0.5^2)$ \\
\hline
\multicolumn{4}{l}{\textit{Curve intercepts}} \\
$\beta^{(1)}$ & Fidelity intercept & $\mathbb{R}$ & $\mathcal{N}(0.5,\,0.5^2)$ \\
$\beta^{(2)}$ & Entropy intercept  & $\mathbb{R}$ & $\mathcal{N}(1,\,2^2)$ \\
$\beta^{(3)}$ & Magic intercept    & $\mathbb{R}$ & $\mathcal{N}(1,\,2^2)$ \\
\hline
\multicolumn{4}{l}{\textit{GP marginal standard deviation}} \\
$\sigma^{(1)}$ & Fidelity GP amplitude & $(10^{-3},\,\infty)$ & $\mathcal{N}^+(0,\,0.5^2)$ \\
$\sigma^{(2)}$ & Entropy GP amplitude  & $(10^{-3},\,\infty)$ & $\mathcal{N}^+(0,\,2^2)$ \\
$\sigma^{(3)}$ & Magic GP amplitude    & $(10^{-3},\,\infty)$ & $\mathcal{N}^+(0,\,2^2)$ \\
\hline
\multicolumn{4}{l}{\textit{GP lengthscale}} \\
$\ell^{(1)}$ & Fidelity GP lengthscale & $(0,\,\infty)$ & $\mathrm{LogNormal}(0,\,1)$ \\
$\ell^{(2)}$ & Entropy GP lengthscale  & $(0,\,\infty)$ & $\mathrm{LogNormal}(0,\,1)$ \\
$\ell^{(3)}$ & Magic GP lengthscale    & $(0,\,\infty)$ & $\mathrm{LogNormal}(0,\,1)$ \\
\hline
\multicolumn{4}{l}{\textit{Excess (unmodeled) noise}} \\
$\eta^{(1)}$ & Fidelity nugget (std.\ dev.) & $(0,\,\infty)$ & $\mathcal{N}^+(0,\,0.5^2)$ \\
$\eta^{(2)}$ & Entropy nugget (std.\ dev.)  & $(0,\,\infty)$ & $\mathcal{N}^+(0,\,1)$ \\
$\eta^{(3)}$ & Magic nugget (std.\ dev.)    & $(0,\,\infty)$ & $\mathcal{N}^+(0,\,1)$ \\
\hline
\end{tabular}
\end{table*}

\subsection{Numerics for Model Normality Assumption}\label{normality_numerics}

In \cref{GP_details}, we explain that our model assumes observed data are normally distributed about the GP mean at each scaled data point. We state that we make this assumption for modeling simplicity. Here, we provide some numerical evidence that this assumption approximately holds.

\begin{figure*}[t]
    \centering
    \begin{subfigure}[b]{0.3\textwidth}
        \centering
        \includegraphics[width=\textwidth]{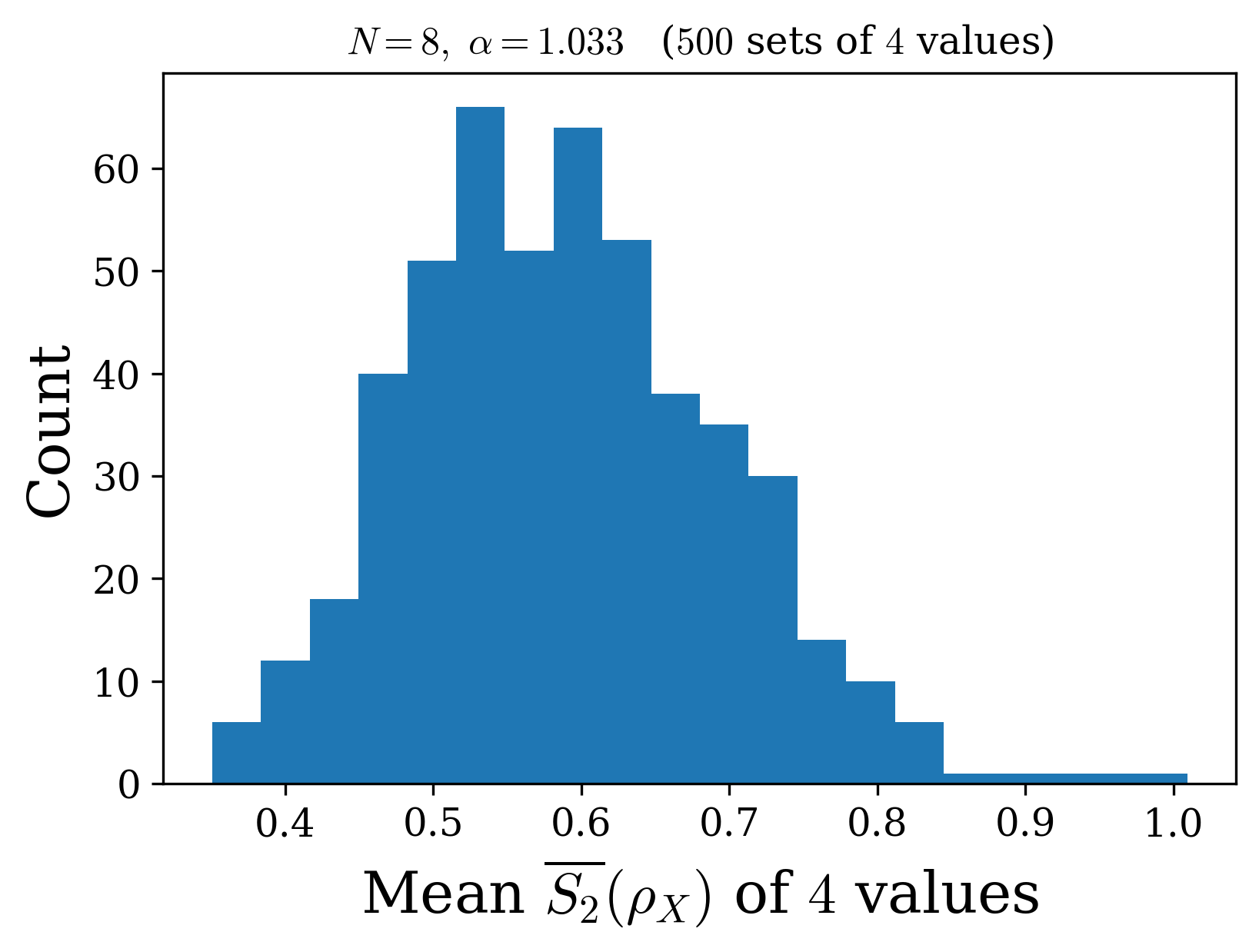}
        \caption{Histogram of $\overline{S_2}(\rho_X)$ values for $N = 8, \alpha = 1.033$, generated from $500$ samples of $4$ values in a set of $1{,}000$ total values.}
        \label{fig:S2_N8_normal}
    \end{subfigure}
    \hfill
    \begin{subfigure}[b]{0.3\textwidth}
        \centering
        \includegraphics[width=\textwidth]{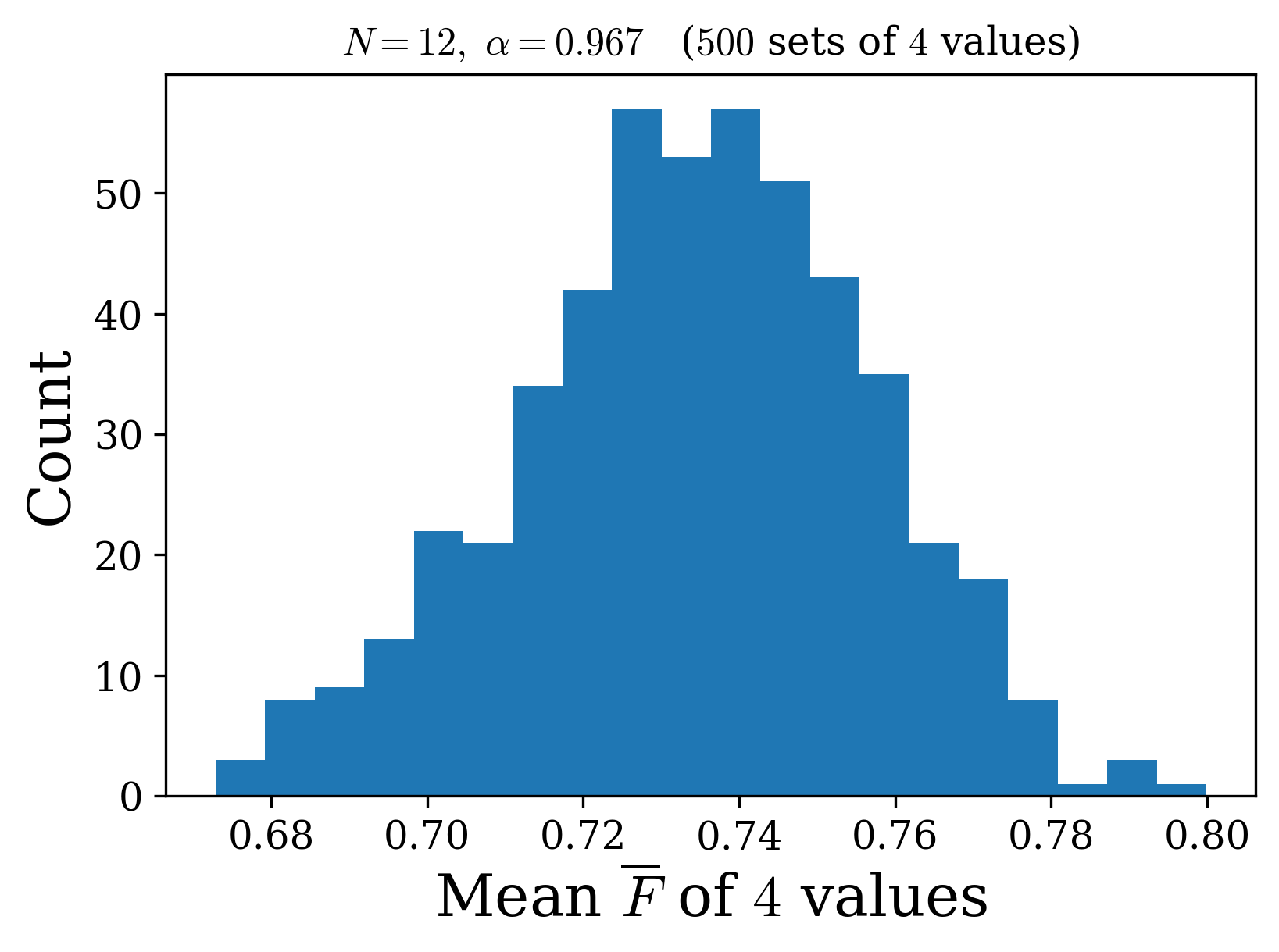}
        \caption{Histogram of $\overline{F}$ values for $N = 12, \alpha = 0.967$, generated from $500$ samples of $4$ values in a set of $1{,}000$ total values.}
        \label{fig:Fbar_N12_normal}
    \end{subfigure}
    \hfill
    \begin{subfigure}[b]{0.3\textwidth}
        \centering
        \includegraphics[width=\textwidth]{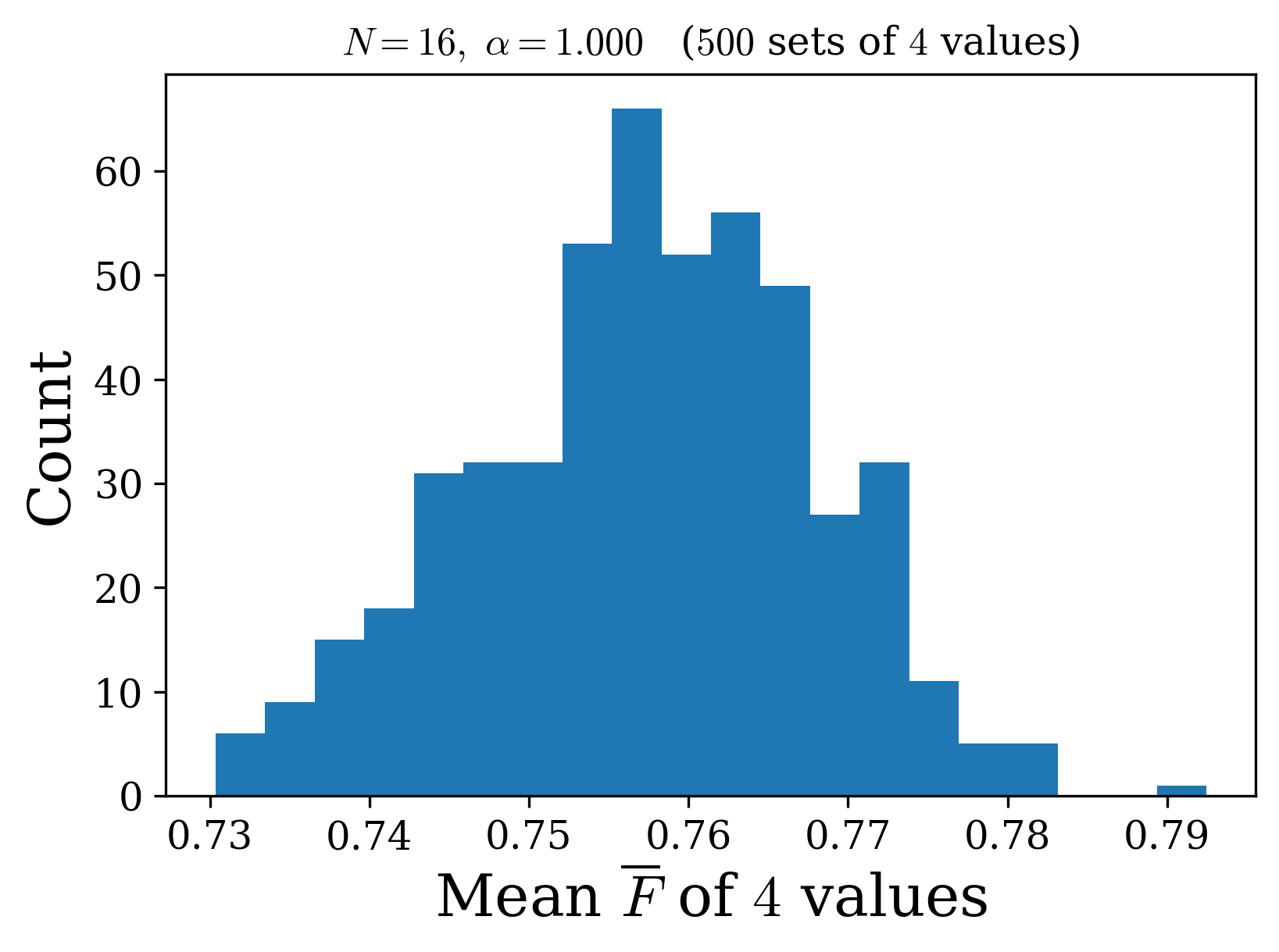}
        \caption{Histogram of $\overline{S_2}(\rho_X)$ values for $N = 16, \alpha = 1.000$, generated from $500$ samples of $4$ values in a set of $200$ total values.}
        \label{fig:Fbar_N16_normal}
    \end{subfigure}
    \caption{Histograms of average order parameters for various $(N, \alpha)$ pairs, from simulated data.}
    \label{fig:histogram_plots_normal}
\end{figure*}

We simulate circuits of $1{,}000$ encoding unitaries for $N = 8$, $1{,}000$ for $N=12$, and $200$ for $N=16$, with the $10$ $\alpha$ values excluding $0, \pi$. For each, we compute the true $F$ and $S_2(\rho_X)$. Then, for each $(N, \alpha)$ pair, we randomly select $4$ values and average them, matching what we implemented experimentally. Repeating this averaging $500$ times for each $N$ gives histograms of $F$ and $S_2(\rho_X)$ whose distributions approximate the expected distributions of our experimental data. We plot some representative histograms in \cref{fig:histogram_plots_normal}. Each looks roughly normally distributed, indicating our assumption is reasonably accurate in addition to simple for modeling purposes.

\section{Appendix - Post-selection Frequency}\label{appendix_postselection}

It is reasonable to wonder what shot overhead this protocol has, even within the EPP. In other words, for a given $(N, \alpha)$, what is the probability that the ancilla measurement returns $\ket{0_{\overline{X}}}$? We first derive this analytically for the model in \cref{theory} then compare with experimental data. We see that the post-selection probabilities falls exponentially in $N$, and the experimental data aligns closely with the theory.

For generic $\mathcal{E}$ and $\mathcal{U}$, the post-selection probability $p_p$ is
\begin{widetext}
\begin{equation}
    p_p = \Tr\left[\left(\mathbb{I}_X \otimes \ket{0_{\overline{X}}}\bra{0_{\overline{X}}}\right)\mathcal{U}^\dagger\mathcal{E}^{\otimes N}\mathcal{U}\left(\ket{0_X}\bra{0_X} \otimes \ket{0_{\overline{X}}}\bra{0_{\overline{X}}}\right) \right].
\end{equation}

Expanding with $\mathcal{E}(\cdot) = \sum\limits_i K_i(\cdot)K_i^\dagger$ gives:
\begin{equation}
    p_p = \Tr\left[\left(\mathbb{I}_X \otimes \ket{0_{\overline{X}}}\bra{0_{\overline{X}}}\right)\sum\limits_i U^\dagger K_i U \left(\ket{0_X}\bra{0_X} \otimes \ket{0_{\overline{X}}}\bra{0_{\overline{X}}}\right)U^\dagger K_i^\dagger U \right].
\end{equation}
Using the swap trick yields:
\begin{equation}\label{p_p_equation}
    p_p = \sum\limits_i \Tr \left[S U^{\dagger \otimes 2} K_i \otimes K_i^\dagger U^{\otimes 2} \left( \ket{0_X}\bra{0_X} \otimes \ket{0_{\overline{X}}}\bra{0_{\overline{X}}} \right) \otimes \left( \mathbb{I}_X \otimes \ket{0_{\overline{X}}}\bra{0_{\overline{X}}} \right) \right],
\end{equation}
where $S$ is the swap between $\mathcal{H}_X \otimes \mathcal{H}_{\overline{X}}$ and a second copy of the Hilbert space, $\mathcal{H}_{X'} \otimes \mathcal{H}_{\overline{X'}}$. Now, to get $\overline{p_p}$, we can take the Haar average (see, e.g.~\cite{haar_intro_mele}), which projects onto $\frac{1+S}{2}$ and $\frac{1-S}{2}$:
\begin{equation}
    \mathbb{E}_U \left(U^{\dagger \otimes 2} K_i \otimes K_i^\dagger U^{\otimes 2}\right) = \frac{1}{2}\left[|\Tr(K_i)|^2 + \Tr(K_i K_i^\dagger) \right]\frac{1+S}{d(d+1)} + \frac{1}{2}\left[\Tr(K_i)^2 - \Tr(K_i K_i^\dagger) \right]\frac{1-S}{d(d-1)},
\end{equation}
\end{widetext}
where $d \equiv 2^N$. In the model, $K_i = \prod\limits_j e^{-iZ_j\alpha/2}$, meaning $\Tr(K_i K_i^\dagger) = \Tr(\mathbb{I}) = d$ and $K_i = \prod_j\cos(\frac{\alpha}{2})\mathbb{I} + i\sin(\frac{\alpha}{2})Z_j$, so that $\Tr(K_i)^2 = d^2 \cos^{2N}(\frac{\alpha}{2})$.

\begin{figure}[h]
    \centering
    \includegraphics[width=\columnwidth]{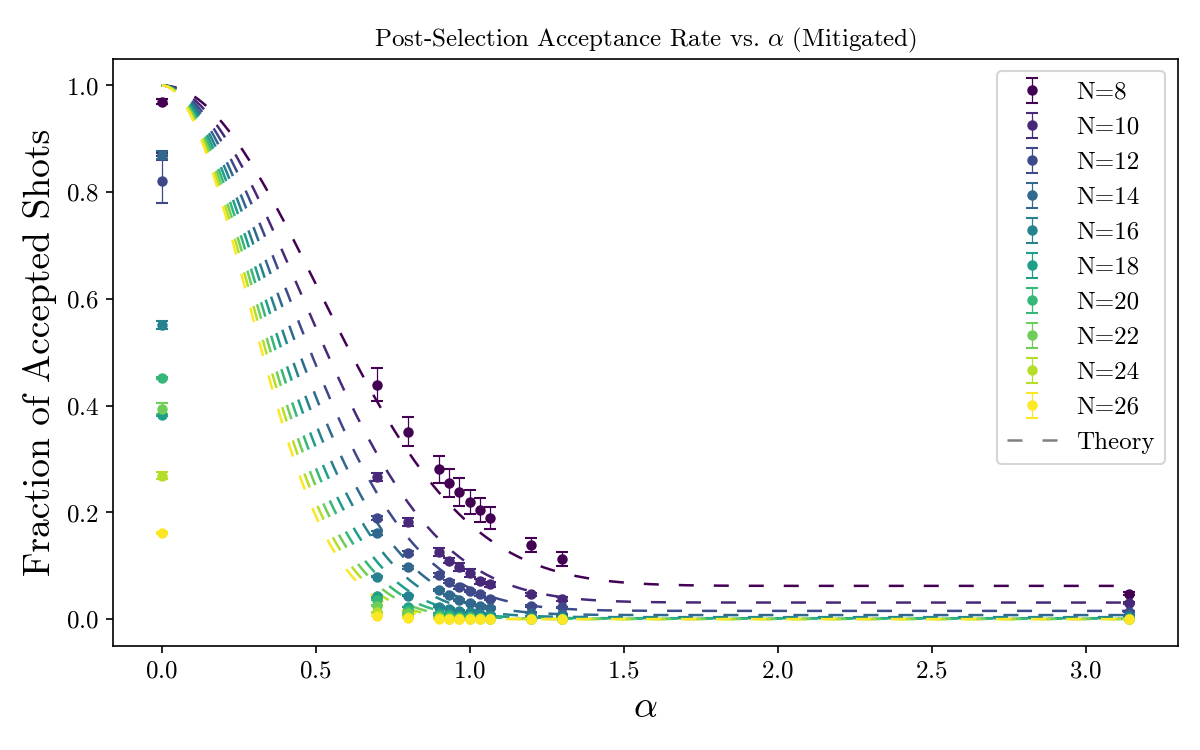}
    \caption{\textbf{Solid lines)} Experimental frequency of observing $\ket{0_{\overline{X}}}$ in ancilla measurement \textbf{Dashed)} Theoretical probability of same outcome for model in \cref{theory}.}
    \label{fig:postselection_probs}
\end{figure}

Plugging this back into \cref{p_p_equation} and simplifying gives:
\begin{equation}\label{p_p_average}
    \overline{p_p}(N, \alpha) = \frac{(d^{3/2} - 1) + (d^2-d^{3/2})\cos^{2N}(\frac{\alpha}{2})}{d^2 - 1}.
\end{equation}
The dashed lines in \cref{fig:postselection_probs} plot this function for the $N$ we experimentally run fidelity experiments over. The experimental and theoretical probabilities closely match.

\bibliographystyle{unsrt}
\bibliography{refs}

\end{document}